\documentclass[twocolumn]{aastex61}

\begin{document}

\newcommand{\kms}{km~s$^{-1}$}  	\newcommand{\mas}{mas~yr$^{-1}$}
\newcommand{\msun}{$M_{\odot}$} 	\newcommand{\rsun}{$R_{\odot}$} 
\newcommand{\teff}{$T_{\rm eff}$}	\newcommand{\logg}{$\log{g}$} 
\newcommand{\vsini}{$v\sin{i}$}

\accepted{March 17, 2017}

\title{The Physical Nature of Subdwarf A Stars: White Dwarf Impostors}

\author{Warren R.\ Brown} \affiliation{Smithsonian Astrophysical Observatory, 60 
Garden St, Cambridge, MA 02138 USA}

\author{Mukremin Kilic} \affiliation{Homer L. Dodge Department of Physics and 
Astronomy, University of Oklahoma, 440 W. Brooks St., Norman, OK, 73019 USA}

\author{A.\ Gianninas} \affiliation{Homer L. Dodge Department of Physics and 
Astronomy, University of Oklahoma, 440 W. Brooks St., Norman, OK, 73019 USA}

\email{wbrown@cfa.harvard.edu, kilic@ou.edu, alexg@nhn.ou.edu}

\shorttitle{ Subdwarf A Stars: White Dwarf Imposters }
\shortauthors{Brown et al.}

\begin{abstract}

	We address the physical nature of subdwarf A-type (sdA) stars and their 
possible link to extremely low mass (ELM) white dwarfs (WDs). The two classes of 
objects are confused in low-resolution spectroscopy. However, colors and proper 
motions indicate that sdA stars are cooler and more luminous, and thus larger in 
radius, than published ELM WDs.  We demonstrate that surface gravities derived from 
pure hydrogen models suffer a systematic $\sim$1~dex error for sdA stars, likely 
explained by metal line blanketing below 9000~K.  A detailed study of five eclipsing 
binaries with radial velocity orbital solutions and infrared excess establishes that 
these sdA stars are metal-poor $\simeq$1.2~\msun\ main sequence stars with 
$\simeq$0.8~\msun\ companions.  While WDs must exist at sdA temperatures, only 
$\sim$1\% of a magnitude-limited sdA sample should be ELM WDs.  We conclude that the 
majority of sdA stars are metal-poor A--F type stars in the halo, and that recently 
discovered pulsating ELM WD-like stars with no obvious radial velocity variations 
may be SX~Phe variables, not pulsating WDs.

\end{abstract}


\section{INTRODUCTION}

	Since the pioneering work of \citet{payne25}, astronomers have deduced 
the physical properties of stars through the combination of spectroscopy and stellar 
atmosphere models.  Different spectral features are sensitive to different stellar 
atmosphere parameters.  The modern approach is to compare an observed stellar 
spectrum to a grid of synthetic spectra calculated over a wide range of parameters.  
The most important parameters are effective temperature \teff, surface gravity 
\logg, and metal abundance [M/H].  Once the stellar parameters are constrained, a 
comparison with evolutionary tracks yields a physical interpretation of the star.  
Consider a spectral A-type, \teff\ = 8000~K star, which is the focus of this paper.  
If the star has \logg\ = 8, it is a normal hydrogen-atmosphere WD; if the star has 
\logg\ = 4, it is a normal main sequence A-type star.  But if the star has \logg\ = 
6, it is something else.

	In 2010 we began the ELM Survey, a spectroscopic survey targeting extremely 
low mass (ELM) WD candidates in the range $5\lesssim \log{g} \lesssim7$ and 8000~K 
$\lesssim T_{\rm eff} \lesssim$ 22,000~K \citep{kilic10, brown10c}.  ELM WDs are 
interesting because they form in ultra-compact binaries; the Universe is not old 
enough to make them through single-star evolution \citep{webbink84, iben90, 
marsh95}.  Indeed, we have found and published 76 ELM WDs in short-period binaries 
over the past seven years \citep{kilic11a, kilic12a, brown12a, brown13a, brown16a, 
gianninas15}.  The median binary orbital period is 6 hrs, which means that half of 
the observed sample will merge within a Hubble time.  The merger rate implies that 
most ELM WD binaries will not undergo stable mass transfer or explode as supernovae, 
but will undergo unstable mass transfer and merge into single massive WDs 
\citep{brown16b}.

	A recent analysis of Sloan Digital Sky Survey (SDSS) spectra has complicated 
the picture of ELM WDs.  \citet{kepler15, kepler16} fit pure hydrogen and pure 
helium atmosphere models to the spectra of candidate WDs in SDSS, and find thousands 
of \logg\ $\sim$ 6 objects at \teff\ $<$ 9000~K temperatures.  The number of these 
\logg\ $\sim$ 6 objects is too large, by two orders-of-magnitude, to be explained as 
the cooler cousins of ELM WDs found in the ELM Survey.  Yet their surface gravities 
are too great to be explained as metal-poor main sequence stars.  
	Arguments both for and against the ELM WD interpretation appear in
conference proceedings \citep{pelisoli17, hermes17}.  Kepler and collaborators refer
to these objects as subdwarf A (sdA) spectral-type stars, and their physical
interpretation is unclear.

	Perhaps sdA stars are thermally bloated WDs, like the class of EL~CVn 
binaries discovered by \citet{maxted13, maxted14}.  However thermally bloated WDs, 
which are believed to have recently had their envelopes stripped by companions, are 
10 to 100 times lower in luminosity, and 3 to 10 times smaller in radius, than their 
more massive companions \citep[e.g.,][]{carter11, maxted14, rappaport15}. sdA stars 
differ because they {\it are} the primaries:  sdA stars dominate the light of their 
systems, and sdA stars in the eclipsing binaries presented here have comparable 
radii to their companions.

	In this paper we address two questions at the base of the sdA mystery.
1) Are we measuring \teff\ and \logg\ correctly?  2) Are we interpreting \teff\ and 
\logg\ correctly?

	For the published ELM Survey sample, those objects with $10,000 < T_{\rm 
eff} < 20,000$~K are almost certainly ELM WDs.  The spectra of these stars are 
dominated by hydrogen Balmer lines, lines shaped by Stark broadening and thus 
sensitive to temperature and gravity at the temperature of B-type stars.  The 
detached eclipsing binary J0651 proves the point:  its radial velocity orbital 
solution, eclipse light curve, and spectroscopic stellar atmosphere parameters all 
demonstrate that this is a $M=0.25$ \msun, $R=0.037$ \rsun\ ELM WD orbiting another 
WD \citep{brown11b, hermes12c}. The same is also true for two additional eclipsing 
binary systems, GALEX J1717+6757 \citep{vennes11} and SDSS J0751-0141 
\citep{kilic14}.

	The situation changes between A0 and F0 spectral types at cooler 
temperatures (10,000~K to 7500~K) where the sdA stars are found.  Metal line 
blanketing is important below 10,000~K, and the Balmer lines become insensitive to 
temperature \citep{strom69}.  Most metals have ionization potentials of 5--8 eV that 
are lower than H (13.5 eV), and so, despite being rare, metals can contribute a 
significant fraction of electron pressure in the atmosphere of A-type stars 
\citep{stromgren69}.  While we would expect metallicities of A-type main sequence 
stars to be low -- such stars would be at $\sim$10 kpc depths in the halo in our 
magnitude-limited survey -- about one-sixth of A-type stars are chemically peculiar 
and enhanced in metals \citep{abt95}.  Helium is a particularly problematic element 
because its abundance is difficult to measure in A-type spectra.  The ionization 
potential of He is higher than H, which means He does not add to electron pressure 
or opacity at A-type temperatures, but it does add to the weight of the atmosphere 
\citep{stromgren44}.


	Spectroscopic analysis of the metal-rich WD GD~362 provides a telling 
example.  Based on low-resolution spectra and pure hydrogen atmosphere models, 
\citet{gianninas04} derive an erroneously high $\log{g}=$ 9.1 for this star.  
Follow-up high resolution spectroscopy and parallax observations demonstrate that 
GD~362 has a helium-dominated atmosphere and $\log{g}=$ 8.2 \citep{zuckerman07, 
kilic08}.

	Yet another complication is that the classic instability strip crosses the 
main sequence at A- and F-type temperatures.  About one-third of A-type stars are 
observed to be $\delta$ Scuti variables, or SX Phe variables if they are metal-poor 
\citep{breger00}.  Fitting models to spectra is also complicated by the fact that 
the blackbody peak moves across the Balmer discontinuity at A- and F-type 
temperatures.  Thus there are many reasons to be cautious about deriving the 
parameters of sdA stars.


	We constrain the physical nature of sdA stars with a three-pronged approach.  
First, we compare the observed distributions of WDs and sdA stars.  Second, we fit 
pure hydrogen models to synthetic main sequence spectra to quantify systematic error 
in the derived temperature and gravity.  We also compare alternative SEGUE Stellar 
Parameter Pipeline fits to sdA spectra.  Third, we study in detail a sample of 11 
sdA stars suspected of being eclipsing binaries (Kepler 2015, private communication) 
and 11 previously unpublished ELM WD candidates with sdA-like temperatures.  Most of 
the stars in this sample have significant radial velocity variability; 6 objects 
have well-measured eclipse light curves; 5 objects have significant infrared excess. 
A joint constraint on the observations suggests that sdA stars are metal-poor A-F 
type main sequence stars with late-type companions.

\section{SURVEYS}

	We begin by introducing the published samples of sdA stars and ELM WDs. A 
comparison of basic observational parameters shows that sdA stars differ 
systematically from ELM WDs in color (temperature) and in reduced proper motion 
(luminosity).

\subsection{sdA Star Sample}

\begin{figure} 
 \includegraphics[width=3.5in]{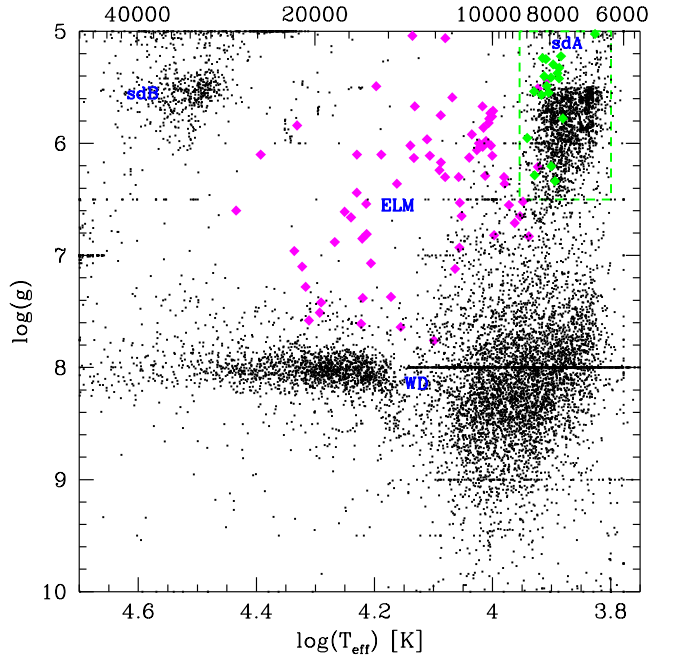}
 \caption{ \label{fig:sda}
	Distribution of \citet{kepler15, kepler16} \teff\ and \logg\ derived from 
pure hydrogen and pure helium model fits to SDSS spectra (black dots).  Clouds of 
normal sdB stars and WDs are labeled; the cloud of sdA stars is marked by the green 
dashed rectangle.  Published ELM WD binaries are over-plotted as magenta diamonds.  
Green diamonds mark the sdA stars studied here.
	}
\end{figure}

	The sample of sdA stars comes from the SDSS WD catalogs of \citet{kepler15, 
kepler16}.  SDSS acquires spectra using a complicated target selection that depends 
on magnitude, color, proper motion, and position as summarized by \citet{kepler15, 
kepler16}.  \citet{kepler15, kepler16} fit pure hydrogen and pure helium atmosphere 
models to these SDSS spectra to identify WDs.  Because of the underlying target 
selection, the catalog of SDSS WDs is incomplete in magnitude, color, proper motion, 
and position.

	Figure \ref{fig:sda} plots the distribution of \teff\ and \logg\ derived 
from the pure hydrogen and pure helium model atmosphere fits \citep{kepler15, 
kepler16}.  We draw Figure \ref{fig:sda} like an H-R diagram, with temperature 
increasing to the left and gravity decreasing (and thus luminosity increasing) to 
the top.  The largest cloud of objects are normal DA and DB WDs scattered around 
\logg\ =8.  Horizontal lines at fixed \logg\ are DC and DZ WDs for which only 
temperature can be measured.  There is also a small cloud of objects with \logg\ 
=5.5 around 30,000~K which are the helium-burning subdwarf B stars.  Low mass WD 
binaries published in the ELM Survey, described below, are marked by magenta 
diamonds.

	Figure \ref{fig:sda} shows a significant, and unexpected, cloud of objects 
with $\log{g} \sim 6$ at cool \teff\ $<9000$ K temperatures:  the sdA stars.  Given 
the observed distribution, we define sdA stars as having 6500~K $<$ \teff\ $<$ 
9000~K, $5 < \log{g} < 6.5$ (green dashed box) for the purposes of this paper.  
There are about 2600 sdA stars in this sample.  Some of these sdA stars are 
binaries, systems that allow us to place physical constraints on the nature of sdA 
stars.  The green diamonds mark the subset of sdA stars that we study in greater 
detail below.

\subsection{ELM WD Sample}

\begin{figure} 
 \includegraphics[width=3.5in]{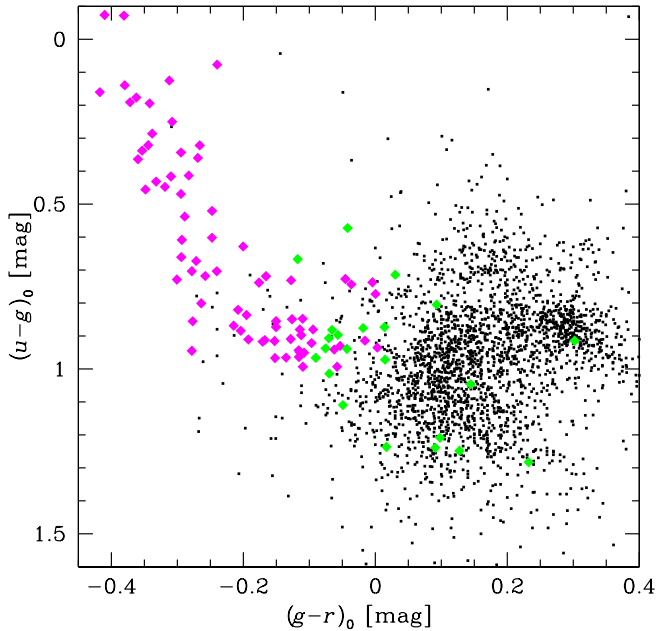}
 \caption{ \label{fig:sda2}
	De-reddened $(g-r)$ versus $(u-g)$ color-color plot for the sdA stars (black 
dots) inside the green dashed box in Fig.\ \ref{fig:sda}, and published ELM WDs 
(magenta diamonds).  Green diamonds mark the sdA stars studied here.
	}
\end{figure}

	The sample of ELM WDs comes from the ELM Survey \citep{kilic10, kilic11a, 
kilic12a, brown10c, brown12a, brown13a, brown16a, gianninas15}.  The ELM Survey is a 
targeted spectroscopic survey of low mass WD candidates selected by color.  The 
sample is complete in magnitude $15 < g < 20$ over the entire SDSS imaging 
footprint; stars with all proper motions are observed.  We focus our attention on 
the 76 published ELM WDs found in single-lined spectroscopic binaries, objects for 
which we have physical constraints on the nature of the stars.

	Figure \ref{fig:sda2} plots the distribution of de-reddened $(g-r)_0$ and 
$(u-g)_0$ color for the ELM WD binaries (magenta diamonds) and the sdA stars (black 
dots, those inside the green dashed box from Figure \ref{fig:sda}).  We draw the 
Figure like an H-R diagram again, with temperature increasing to the left.  The band 
of ELM WD binaries reflects the ELM Survey color selection:  we target stars with 
mid- to late-B type colors, so that is what we find.

	The cloud of sdA stars in Figure \ref{fig:sda2} is systematically redder 
than the ELM WDs.  The difference in $(g-r)_0$ color indicates that sdA stars are 
systematically cooler than published ELM WDs.  The green diamonds in Figure 
\ref{fig:sda2} again mark the subset of sdA stars that we study in greater detail 
below.

\begin{figure} 
 \includegraphics[width=3.5in]{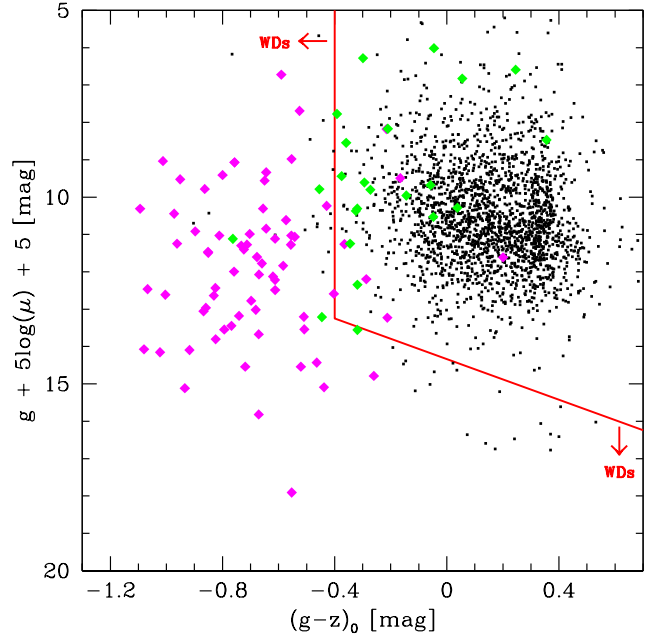}
 \caption{ \label{fig:rpm}
	Reduced proper motion versus de-reddened $(g-z)$ color for the sdA stars 
(black dots) inside the green dashed box in Fig.\ \ref{fig:sda}, and published ELM 
WDs (magenta diamonds).  Solid red line marks the division between SDSS WDs (lower 
left) and quasars (upper right) seen by \citet{fusillo15}.  Green diamonds mark the 
sdA stars studied here.
	}
\end{figure}

\subsection{Reduced Proper Motion Comparison}

	Reduced proper motion is a classic tool for separating stars at the same 
temperature by their luminosity (e.g. subdwarfs, dwarfs, and giants).  
\citet{fusillo15} apply this technique to the SDSS photometric catalog and find a 
clean separation between main sequence stars and WDs using a color-RPM diagram 
\citep[see also][]{kilic06}. We copy their approach for sdA stars and ELM WDs.

	We use proper motions, $\mu$, obtained from the HSOY catalog 
\citep{altmann17}, a new proper motion catalog that uses {\it Gaia} Data Release 1 
positions for its final epoch.  Figure \ref{fig:rpm} plots the distribution of 
reduced proper motion, $H_g = g + 5\log(\mu) +5$, versus de-reddened $(g-z)_0$ 
color.
	As before, sdA stars are plotted as black dots, and ELM WDs are plotted as 
magenta diamonds.  The solid red line marks the division between quasars (luminous 
distant objects), which sit in the upper right, and WDs (faint nearby objects), 
which sit below and to the left \citep{fusillo15}.

	\citet{fusillo15} observe WDs with $(g-z)_0 \simeq 0$ colors similar to the 
sdA stars, but the WDs are intrinsically fainter, $H_g>15$ mag, and fall below the 
red line.  Thus Figure \ref{fig:rpm} shows that sdA stars are systematically more 
luminous for their color (temperature) than WDs; 98\% of sdA stars lie above 
the red line in Figure \ref{fig:rpm}.  Conversely, the ELM WDs have luminosities of 
known WDs at comparable temperatures; 92\% of published ELM WDs lie below or 
to the left of the red line in Figure \ref{fig:rpm}.

\section{STELLAR PARAMETERS}

	In the previous section we find that sdA stars have cooler temperatures and 
greater luminosities than published ELM WDs.  The luminosity of a star is related to 
its radius and effective temperature according to the Stephan-Boltzmann law, 
\begin{equation} L = 4 \pi R^2 \sigma T^4 , \end{equation} where $L$ is luminosity, 
$R$ is radius, $\sigma$ is the Stephan-Boltzmann constant, and $T$ is effective 
temperature.  Equation~1 tells us that, for sdA stars to be cooler and more luminous 
than ELM WDs, {\em they must have larger radii.}

	Surface gravity depends on mass $M$ divided by radius squared, 
\begin{equation} \log{g} = 4.438 + \log{ (M / R^2) }, \end{equation} where $g$ is in 
units of cm s$^{-2}$, $M$ is in units of solar mass, and $R$ is units of solar 
radii.  Equation~2 tells us that a typical $\log{g} = 6$, $M = 0.2$ \msun\ ELM WD 
should have a radius of 0.074 \rsun.  Such radii are observed for ELM WDs 
\citep{hermes14}.

	An sdA star that has the {\em same surface gravity but a larger radius} than 
an ELM WD must have a larger mass.  Quantitatively, a 3 times larger radius requires 
a 9 times larger mass, or 1.8 \msun\, to keep surface gravity constant.  1.8 \msun\
is the mass of an A-type main-sequence star, but $\log{g}=6$ is not the correct 
gravity for a main-sequence star.

	One explanation is that sdA surface gravities derived from pure hydrogen 
models are incorrect.  Metal line blanketing, combined with the insensitivity of the 
hydrogen Balmer lines at these temperatures, may cause the pure hydrogen models to 
yield incorrect parameters.  We test this hypothesis in two ways.

\subsection{Theoretical Test}

\begin{figure} 
 \includegraphics[width=3.5in]{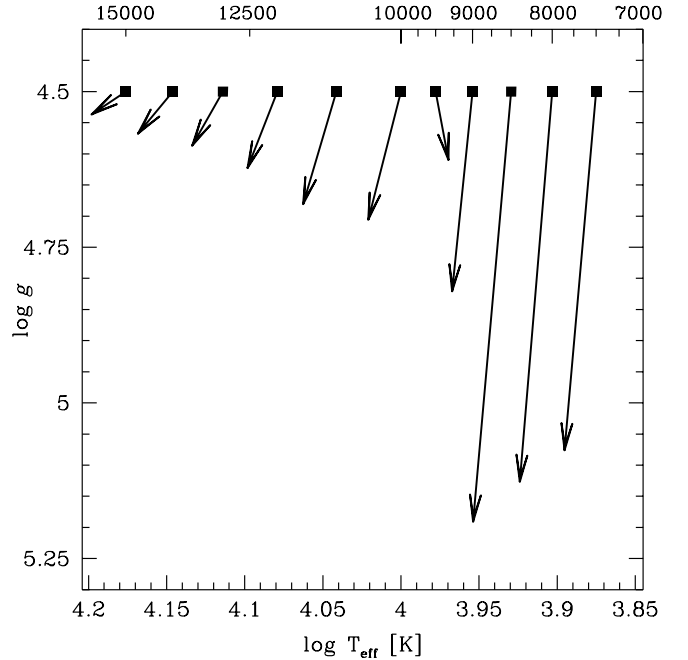}
 \caption{ \label{fig:dtg}
	Systematic error of pure hydrogen model fits (arrows) to synthetic main 
sequence spectra with known parameters (black squares).  For [M/H]=$-1$, shown
here, the systematic error in \logg\ is about 0.6 dex below 9000~K.
	}
\end{figure}

	First, we fit our pure hydrogen model atmospheres to synthetic spectra of 
main sequence B-, A-, and F0-type stars.  We generate synthetic spectra using ATLAS9 
ODFNEW model atmosphere grids \citep{castelli04, castelli97} convolved to the 1 \AA\ 
spectral resolution of our MMT Blue Channel Spectrograph data.  We fix \logg=4.5 and 
step \teff\ from 7500~K to 15,000~K at solar metallicity and at one-tenth solar 
metallicity ([M/H]=$-1$).  We also set \vsini=100 \kms\ appropriate for main 
sequence stars of these temperatures \citep{abt02}.  Setting \vsini=0 \kms\ changes 
the results very little, by less than 1\%.

	We then fit the synthetic spectra to the \citet{gianninas11, gianninas14b} 
grid of pure hydrogen atmosphere models used for the ELM Survey.  
\citet{tremblay13, tremblay15} show that parameters obtained from 1D stellar 
atmosphere fits must be corrected for 3D convection effects, and provide 1D to 3D 
corrections for the range 6000~K $<$ \teff\ $<$ 14,000~K and $5 < \log{g} < 9$.  
Only our coolest $<9000$ K synthetic spectra have fitted parameters that rise above 
$\log{g}=5$ and fall in this range.  The predicted corrections would reduce the 
fitted \teff 's by about 5\% and \logg 's by about 0.1 dex.  However, because we do 
not have corrections for most of the measurements, we do not apply any corrections. 
Our immediate interest is making a self-consistent comparison.

	Figure \ref{fig:dtg} presents the results for the [M/H]=$-1$ synthetic
spectra.  We find that \teff\ fit from the pure hydrogen atmosphere models is
systematically too large by about 5\% across all temperatures.  It is possible that
an appropriate 1D to 3D correction could address this.  Surface gravity shows a
similar systematic error above 9000~K, however the error in \logg\ explodes to 0.6
dex at sdA temperatures below 9000~K.  This is many times larger than tabulated 1D
to 3D corrections \citep{tremblay15}.  If we look at the results for the solar
metallicity synthetic spectra, we find that the systematic error in \logg\ is even
greater below 9000~K, nearly 1.0 dex.  We conclude that pure hydrogen models
systematically over-estimate the \logg\ of main sequence spectra at sdA temperatures
below about 9000~K.


\subsection{Empirical Test}

\begin{figure} 
 \includegraphics[width=3.5in]{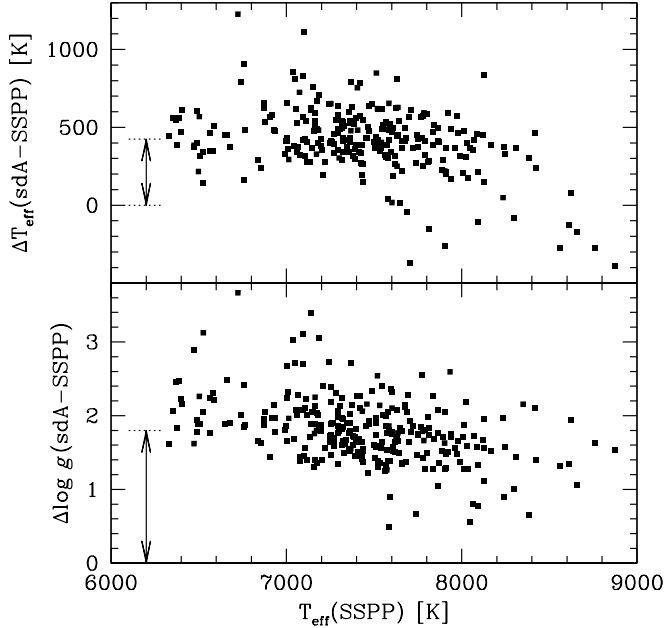}
	\caption{ \label{fig:sspp} Difference between \citet{kepler15, kepler16} 
pure hydrogen model fits and SSPP main sequence fits to sdA stars (black dots) 
inside the green dashed box in Fig.\ \ref{fig:sda}, plotted as a function of SSPP 
effective temperature.
	} 
\end{figure}

	Second, we compare two different fits to the same observed spectra.  We 
cross-match the list of sdA stars \citep{kepler15, kepler16} against the SEGUE 
Stellar Parameter Pipeline \citep[SSPP,][]{lee08, allende08, smolinski11} and find 
287 matches.  This is about 10\% of the sdA list; the other sdA stars are from SDSS 
spectra obtained after Data Release 9 and so are not included in the SSPP catalog.

	Figure \ref{fig:sspp} plots the difference between the pure hydrogen 
atmosphere model fits \citep{kepler15, kepler16} and the SSPP main sequence fits as 
a function of SSPP temperature.  There is a systematic offset:  the median pure 
hydrogen atmosphere \teff\ is 425 K hotter (or 5\% at 8000~K, the same as the 
synthetic tests), and the median \logg\ is 1.8 dex larger than the SSPP main 
sequence model.

	The sdA stars have a median $\log{g} \simeq 4$ in the SSPP catalog; this is 
the surface gravity of a main sequence star.  The sdA stars have a median [Fe/H] = 
$-1.5$ in the SSPP catalog; this is the metallicity of a halo star 
\citep{allende14}.

	Metal-poor stars are less luminous and lower mass than solar-metallicity 
stars of identical temperature.  Consider an 8000~K star.  In the Padova main 
sequence tracks \citep{bressan12} an 8000~K star has 1.9 \msun\ and 2.5 \rsun\ at 
solar metallicity, or 1.2 \msun\ and 1.1 \rsun\ at [M/H] = $-1$.  Thus a metal-poor 
8000~K star is near the main sequence turn-off in the halo; hotter stars may be 
halo blue stragglers.  
	For comparison, an 8000~K ELM WD has 0.18 \msun\ and 0.03 \rsun\ 
\citep{althaus13, istrate16}.

\begin{deluxetable*}{cccccc} 
\tabletypesize{\scriptsize} 
\tablecolumns{6} 
\tablewidth{0pt} 
\tablecaption{Summary of Observations\label{tab:summary}}
\tablehead{
	\colhead{Object} & \colhead{$N_{\rm spec}$} & \colhead{RV Variable} & 
	\colhead{$N_{\rm phot}$} & \colhead{Eclipsing} & \colhead{IR Excess}
}
	\startdata \cutinhead{sdA stars} 
SDSS J074735.035+455420.02 &  6 & \nodata & 351 & \nodata & \nodata \\
SDSS J075017.359+400441.30 & 14 & YES     & 432 & YES     & \nodata \\ 
SDSS J075804.596+475406.10 & 16 & \nodata & 293 & \nodata & \nodata \\
SDSS J080205.902+433228.29 & 10 & \nodata & 407 & YES     & YES \\
SDSS J080313.312+415740.49 & 12 & YES     & 421 & YES     & \nodata \\ 
SDSS J080313.957+383037.73 & 10 & YES     & 457 & \nodata & \nodata \\
SDSS J080540.813+424350.70 &  8 & YES     & 393 & \nodata & \nodata \\ 
SDSS J082328.325+373101.69 &  8 & YES     & 757 & YES     & \nodata \\
SDSS J083238.894+135120.98 & 17 & \nodata & 527 & YES     & YES \\
SDSS J094154.305+370140.92 & 10 & YES     & 373 & \nodata & \nodata \\
SDSS J101132.732+024216.47 &  5 & \nodata & 337 & \nodata & YES \\
	\cutinhead{sdA-like ELM WD candidates}
SDSS J024318.073$-$084749.67 &  8 & YES     & 243 & \nodata & \nodata \\ 
SDSS J042128.509+000213.44   & 27 & YES     & 339 & \nodata & YES \\
SDSS J081115.372$-$021652.75 & 16 & YES     & 147 & \nodata & \nodata \\
SDSS J091157.257+583308.39   &  6 & YES     & 145 & \nodata & \nodata \\
SDSS J140608.922+044302.12   & 11 & YES     & 348 & \nodata & YES \\
SDSS J154102.488+314003.68   & 13 & YES     & 320 & \nodata & \nodata \\
SDSS J160036.831+272117.81   & 20 & YES     & 397 & YES     & YES \\
SDSS J160050.256+051617.27   & 19 & YES     & 345 & \nodata & \nodata \\
SDSS J205219.372$-$033208.71 &  5 & YES     &   0 & \nodata & \nodata \\ 
SDSS J220133.319+242005.63   &  4 & \nodata & 118 & \nodata & \nodata \\
SDSS J221928.482+120418.68   & 27 & YES     & 473 & \nodata & \nodata \\
	\enddata
\end{deluxetable*}

\begin{deluxetable*}{cCCccCCC} 
\tabletypesize{\scriptsize} 
\tablecolumns{8}
\tablewidth{0pt} 
\tablecaption{Stellar Atmosphere Parameters\label{tab:teff}}
\tablehead{
	& \multicolumn{2}{c}{Gianninas} & \multicolumn{2}{c}{Kepler} & \multicolumn{3}{c}{SEGUE Stellar Parameter Pipeline} \\
	\colhead{Object} & \colhead{\teff} & \colhead{\logg} & \colhead{\teff} & \colhead{\logg} & \colhead{\teff} & \colhead{\logg} & \colhead{[Fe/H]} \\
	& (K) & & (K) & & (K) & &
}
	\startdata 
	\cutinhead{sdA stars}
  J0747+45 & 7524 \pm 113 & 5.526 \pm 0.082 &   7546  &  5.749  &    \nodata   &     \nodata     &      \nodata     \\
  J0750+40 & 7998 \pm 124 & 5.335 \pm 0.091 &   8141  &  5.619  & 7524 \pm  46 & 4.229 \pm 0.155 & -0.507 \pm 0.100 \\
  J0758+47 & 7827 \pm 134 & 5.298 \pm 0.154 & \nodata & \nodata & 7360 \pm  65 & 3.549 \pm 0.359 & -0.350 \pm 0.062 \\
  J0802+43 & 7461 \pm 121 & 5.004 \pm 0.144 & \nodata & \nodata & 6769 \pm  64 & 4.380 \pm 0.098 & -0.971 \pm 0.014 \\
  J0803+41 & 8150 \pm 124 & 5.345 \pm 0.075 &   8355  &  5.885  &    \nodata   &     \nodata     &      \nodata     \\
  J0803+38 & 7777 \pm 118 & 6.056 \pm 0.076 &   7646  &  6.303  & 7686 \pm 204 & 4.255 \pm 0.171 & -0.801 \pm 0.240 \\
  J0805+42 & 7870 \pm 120 & 5.909 \pm 0.079 &   7660  &  6.041  & 7815 \pm 176 & 4.211 \pm 0.145 & -1.086 \pm 0.090 \\
  J0823+37 & 7558 \pm 125 & 5.180 \pm 0.142 &   7799  &  5.323  & 7224 \pm  55 & 4.218 \pm 0.124 & -1.010 \pm 0.129 \\
  J0832+13 & 6652 \pm 111 & 4.905 \pm 0.178 &   6786  &  5.359  &    \nodata   &     \nodata     &      \nodata     \\
  J0941+37 & 8337 \pm 130 & 5.982 \pm 0.084 &   7646  &  5.880  & 7906 \pm 183 & 3.825 \pm 0.184 & -1.697 \pm 0.095 \\
  J1011+02 & 8466 \pm 147 & 5.722 \pm 0.128 &   8109  &  5.549  & 7884 \pm  51 & 4.030 \pm 0.210 & -1.220 \pm 0.102 \\
	\cutinhead{sdA-like ELM WD candidates}
J0243$-$08 & 8172 \pm 138 & 5.402 \pm 0.131 & \nodata & \nodata &    \nodata   &     \nodata     &      \nodata     \\
  J0421+00 & 8093 \pm 120 & 5.488 \pm 0.063 & \nodata & \nodata &    \nodata   &     \nodata     &      \nodata     \\
J0811$-$02 & 7709 \pm 116 & 5.326 \pm 0.083 & \nodata & \nodata &    \nodata   &     \nodata     &      \nodata     \\
  J0911+58 & 7961 \pm 154 & 5.413 \pm 0.192 & \nodata & \nodata &    \nodata   &     \nodata     &      \nodata     \\
  J1406+04 & 8119 \pm 131 & 5.249 \pm 0.110 & \nodata & \nodata &    \nodata   &     \nodata     &      \nodata     \\
  J1541+31 & 7891 \pm 119 & 5.296 \pm 0.078 & \nodata & \nodata &    \nodata   &     \nodata     &      \nodata     \\
  J1600+27 & 7910 \pm 138 & 4.775 \pm 0.203 & \nodata & \nodata &    \nodata   &     \nodata     &      \nodata     \\
  J1600+05 & 7762 \pm 115 & 5.374 \pm 0.069 & \nodata & \nodata & 7794 \pm 286 & 4.261 \pm 0.242 & -1.797 \pm 0.114 \\
J2052$-$03 & 8081 \pm 140 & 4.864 \pm 0.169 & \nodata & \nodata &    \nodata   &     \nodata     &      \nodata     \\
  J2201+24 & 8397 \pm 232 & 4.751 \pm 0.511 & \nodata & \nodata &    \nodata   &     \nodata     &      \nodata     \\
  J2219+12 & 8226 \pm 127 & 5.240 \pm 0.073 & \nodata & \nodata &    \nodata   &     \nodata     &      \nodata     \\
	\enddata \tablecomments{Gianninas values are corrected for 3D convection 
effects following \citet{tremblay13, tremblay15}. Kepler values come from 
\citet{kepler15, kepler16}.}

\end{deluxetable*}

\section{NEW OBSERVATIONS}

	To test whether or not sdA stars are indeed metal-poor main sequence stars, 
we study a sub-sample of sdA stars in greater detail.  Our sample is comprised of 11 
sdA stars suspected of being eclipsing binaries (Kepler 2015, private communication) 
and 11 previously unpublished ELM WD candidates that have sdA-like temperatures 
summarized in Table \ref{tab:summary}.  We obtain time-series spectroscopy for all 
22 objects and time-series optical photometry for 21 objects.  We also obtain $JHK$ 
infrared photometry for 6 objects.  We find that 17 objects are radial velocity 
variable, 6 are eclipsing, and 4 of these have significant infrared excess.  All of 
these objects are identified with their full coordinates in Table \ref{tab:summary}.

\subsection{Spectroscopy}

	We obtain time-series spectroscopy for 20 of the 22 objects with the 6.5m 
MMT telescope.  We configured the Blue Channel spectrograph \citep{schmidt89} to 
obtain 1.0 \AA\ resolution spectra, and set exposure times to yield signal-to-noise 
ratios of 10 to 15 in the continuum.  We obtain spectra for the 2 brightest objects 
with the 1.5m Tillinghast telescope at Fred Lawrence Whipple Observatory.  We 
configured the FAST spectrograph \citep{fabricant98} to obtain 1.7 \AA\ resolution 
spectra, and set exposure times to yield similar signal-to-noise ratios.  We obtain 
additional spectra for 6 objects with the 4m Mayall telescope at Kitt Peak National 
Observatory.  We configured the KOSMOS spectrograph \citep{martini14} to obtain 2.0 
\AA\ resolution spectra.  The spectra were mostly acquired in observing runs between 
December 2014 and December 2016.

	All spectra are paired with comparison lamp exposures for accurate 
wavelength calibration.  We measure radial velocities using the RVSAO 
cross-correlation program \citep{kurtz98}.  For each object we typically obtain 
multiple observations on back-to-back nights in different observing runs.  This 
observing pattern allows use to constrain the orbital properties of binaries that 
have periods ranging from hours to days.

\subsection{Stellar Atmosphere Fits}

	We begin by summing the rest-frame spectra and performing stellar atmosphere 
fits to each object.  Our approach is identical to previous ELM Survey papers.  In 
brief, we fit the summed spectra to a grid of pure hydrogen atmosphere models that 
span 4000~K $<$ \teff $<$ 35,000~K and 4.5 $< \log{g} <$ 9.5 \citep{gianninas11, 
gianninas14, gianninas15}.  We apply the \citet{tremblay13, tremblay15} 1D to 3D 
corrections to all of the temperatures and gravities (however we are forced to 
extrapolate the correction for the three objects with $\log{g}<5$).  We present the 
final parameters in Table \ref{tab:teff}, along with the parameters independently 
published by \citet{kepler15, kepler16} and by the SSPP catalog, when available.

	We find good agreement between the values derived from the two pure hydrogen 
models, labeled ``Gianninas'' and ``Kepler'' in Table \ref{tab:teff}.  Clipping one 
outlier (J0941+37), the temperatures differ on average by 6~K with a 
root-mean-square (RMS)  dispersion of $\pm$216~K.  The gravities differ on average 
by 0.20~dex with an RMS dispersion of $\pm$0.23~dex.  The dispersions are consistent 
with the errors summed in quadrature.  Thus we conclude that the two sets of 
hydrogen models give consistent answers.

	The SSPP parameters are systematically different, however, about the same as 
before (Section 3.2).  The SSPP temperatures for these objects are on average 430~K 
cooler than our hydrogen model values, and the gravities are 1.7 dex lower.  
Surface gravity remains the most divergent parameter, and thus the most suspect.  
The offsets are in the same direction, and of similar magnitude, as we found from 
fitting hydrogen models to synthetic main sequence spectra.

\subsection{Radial Velocities}

	Our time-series spectroscopy provides a constraint on the radial velocity 
variability of the 22 objects.  We start with an F-test, to test whether the 
variance of the observed radial velocities exceeds the variance expected from 
measurement errors for an object at rest.  Seventeen objects exhibit significant 
radial velocity variability at better than 99\% confidence, and thus are very likely 
binaries.  The large number of binaries is expected:  half of the objects are 
candidate eclipsing binaries. We identify the velocity variable objects in Table 
\ref{tab:summary}, and provide the individual measurements in the Appendix.

	We calculate orbital elements for the velocity variable objects using the 
same procedure as in previous ELM Survey papers.  In brief, we minimize $\chi^2$ for 
circular orbits using the code of \citet{kenyon86}.  We search for orbital periods 
up to 4 days, which is the maximum time span of our individual observing runs.

\begin{figure} 
 \includegraphics[width=3.5in]{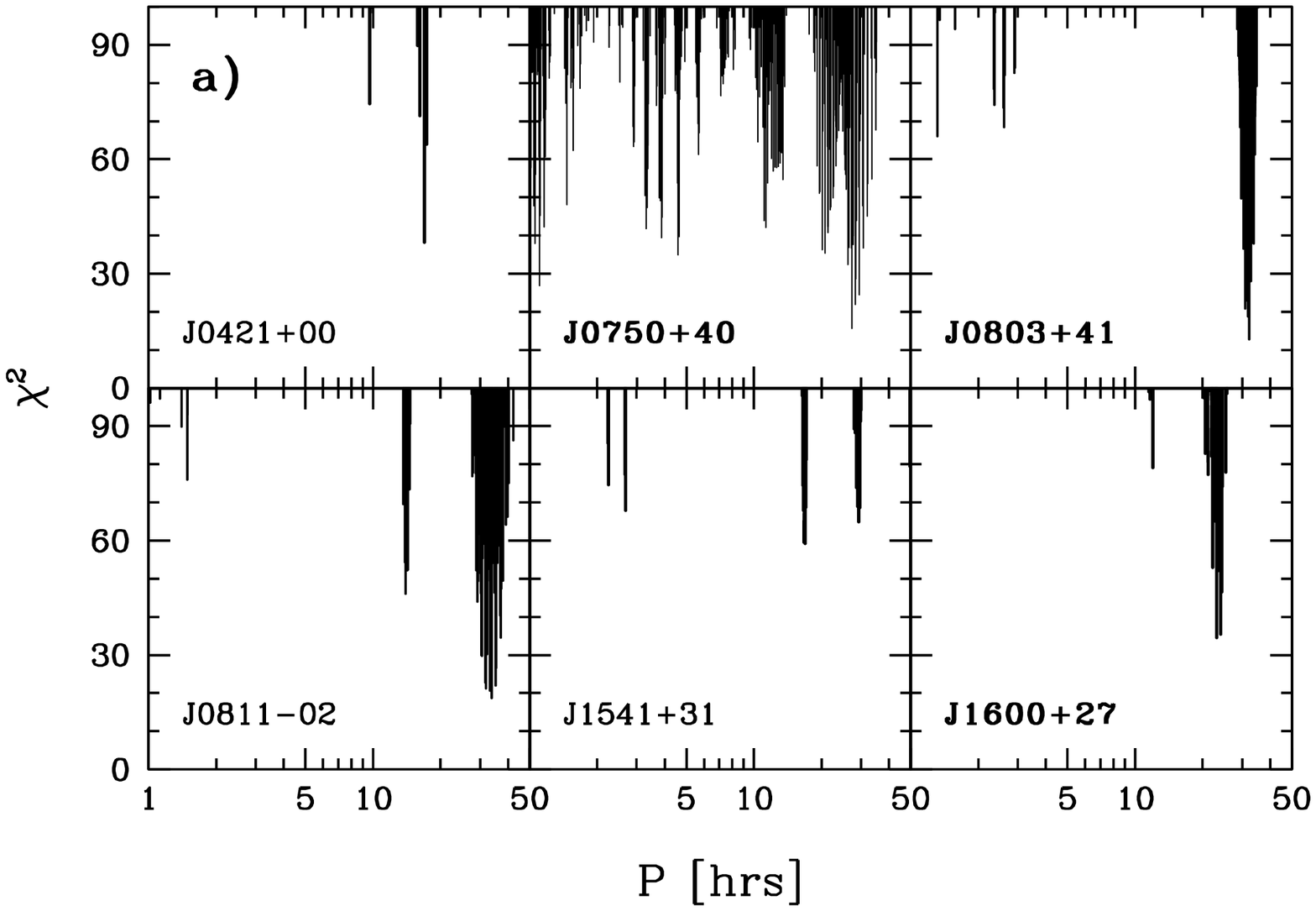} 
 \includegraphics[width=3.5in]{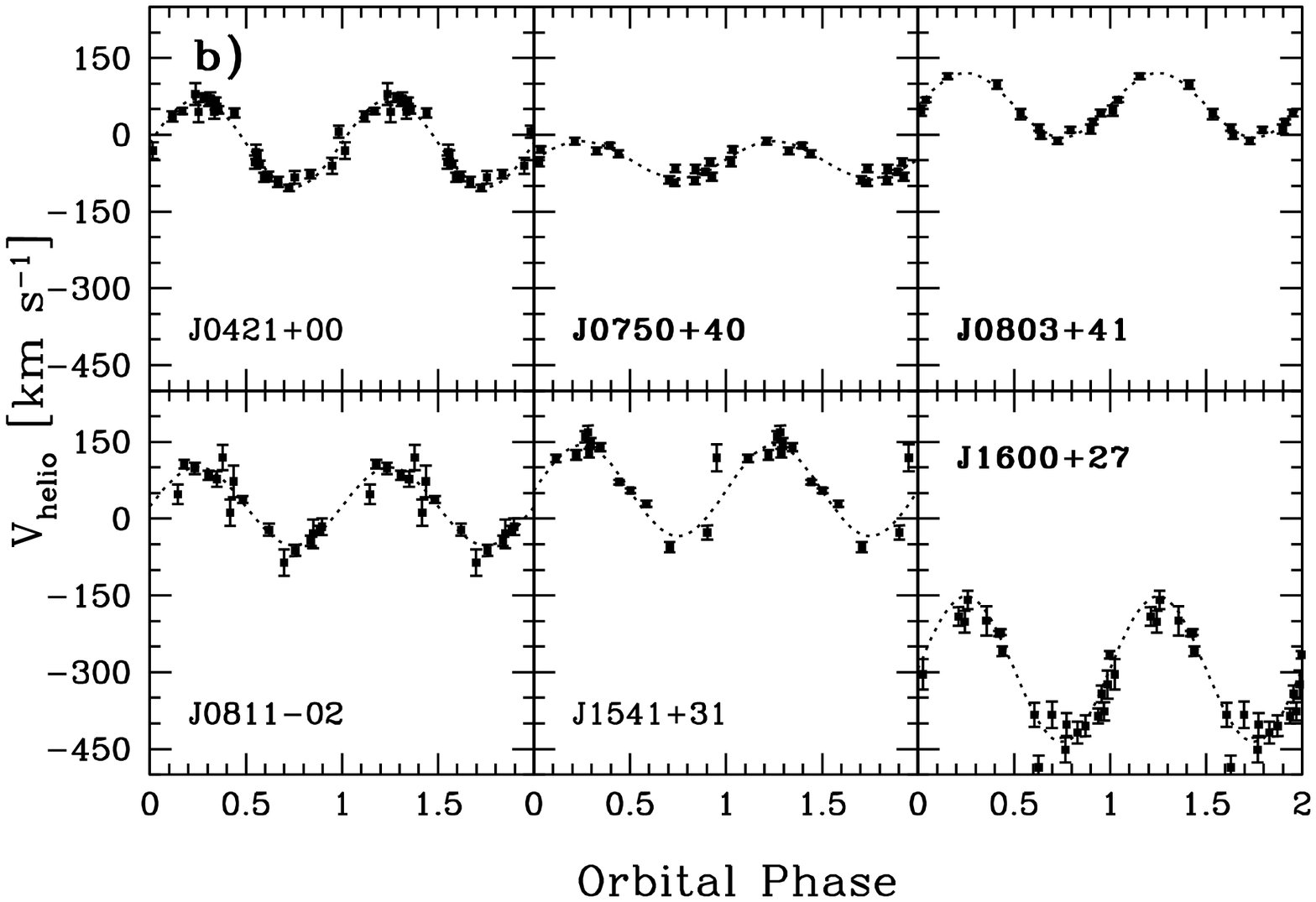}
 \caption{ \label{fig:pdm}
	(a) Periodograms for 6 objects with well-constrained orbital solutions 
(upper panel).  (b) Heliocentric radial velocities phased to the best-fit periods.
	}
\end{figure}

	Six objects have well-constrained orbital parameters; the remaining velocity 
variable objects have insufficient observations or phase coverage to yield unique 
orbital solutions.  We plot the periodograms and the phased radial velocities for 
the well-constrained objects in Figure \ref{fig:pdm}.

	We estimate the significance of possible orbital period aliases for these 
six objects using the $\chi^2$ values.  For normally distributed errors, a $\Delta 
\chi^2=13.3$ with respect to the minimum value corresponds to a 99\% confidence 
interval for 4 degrees of freedom \citep{press92}.  On this basis, three objects 
have no significant period aliases:  J0421+00 is $P=16.94$ hr, J0750+40 is $P=28.35$ 
hr, and J0803+41 is $P=32.27$ hr.  Three other objects have significant but 
uninteresting period aliases:  J0811$-$02 is $P=33.8$ hr with a cloud of aliases 
between 31.7 hr and 35.3 hr, J1541+31 is either $P=16.8$ hr or 29 hr, and J1600+27 
is either $P=23.10$ or 24.08 hr.  We present the orbital parameters for these six 
objects in Table \ref{tab:orbit}.

	Unlike published ELM WD binaries, we find no evidence for short-period or 
high-amplitude velocity variability in any of the 22 sdA-like objects studied here.  
The day-long orbital periods that we do observe are confirmed by light curves.

\subsection{Light Curves}

	We search the Catalina Surveys Data Release 2 \citep{drake09} and find 
time-series $V$-band photometry for 21 of the 22 objects.  Six objects show 
significant eclipses.
	Three of the eclipsing systems have orbital periods in excellent agreement
with their radial velocity solutions.  We adopt the photometric periods for these
systems (J0750+40, J0803+41, and J1600+27).  

\begin{figure} 
 \includegraphics[width=3.5in]{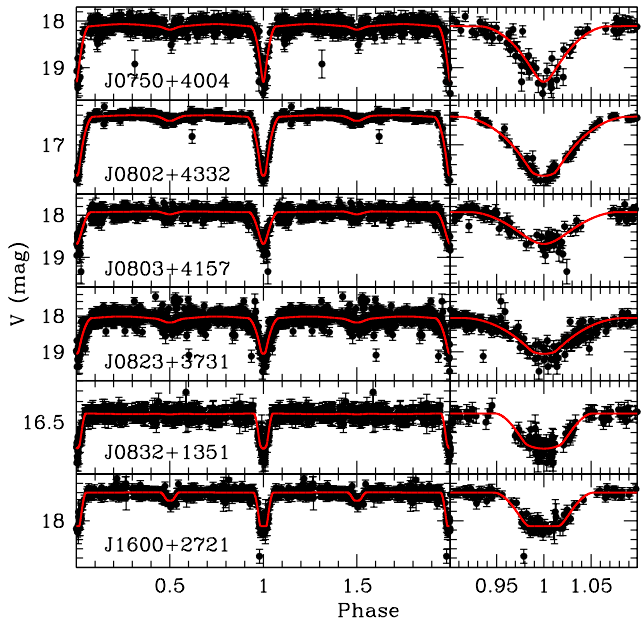} 
 \caption{ \label{fig:lc}
	Phased-folded Catalina $V$-band photometry for 6 eclipsing binaries (black 
points) overplotted with the best-fit light curve models (red lines).
	}
\end{figure}

	Figure \ref{fig:lc} displays the Catalina light curves and our best-fit 
model for the six eclipsing systems.  We model the light curves using JKTEBOP 
\citep{southworth04}.  Based on our initial fits, all six binary systems are 
composed of stars with radii consistent within a factor of two.  Hence, we assume a 
mass ratio of 1 in our light curve fits and use the appropriate gravity and linear 
limb darkening coefficients for $T_{\rm eff}=$ 8000~K, $\log{g}=4$, and solar 
metallicity stars from \citet{claret11}.  We perform 100 Monte Carlo simulations for 
each star to estimate the errors in each parameter.  

	Table \ref{tab:lc} presents our best-fit model parameters for the six 
eclipsing systems, whose periods range from 0.7 to 3.7 days. It is striking that all 
six systems include binary stars with radii that range from 10\% to $\approx$30\% of 
the orbital separation. These stars are too large to be WDs; they are clearly 
main-sequence stars.  Note that the average radius of the primary star in J0802+4332 
is nearly 30\% of the orbital separation, so this radius may be wrong by 5\% 
\citep{north04}.

	The assumed mass ratio, $q=M_2/M_1$, has negligible effect on the radii 
measurements for all but two of the stars, J0802+4332 and J0803+4157.  Yet even for 
$q=0.25$, the radii estimates are larger than 10\% of the orbital separation for all 
primary and secondary stars in all six systems.  Thus our choice of $q$ does not 
affect the conclusion that these stars are too large to be WDs.

	Table \ref{tab:lc} also presents the $V-$band luminosity ratios, $L_2 / 
L_1$, for the eclipsing systems.  The secondary eclipses for these six objects are 
less well defined, compared to the primary eclipses. This results in large 
uncertainties in the luminosity ratios of the two stars in each system.  With the 
exception of J1600+27, the eclipse ratios are consistent with luminosity ratios of 
about ten of percent in the $V-$band.

\begin{deluxetable}{cccCR} 
\tabletypesize{\scriptsize} 
\tablecolumns{5} 
\tablewidth{0pt} 
\tablecaption{Radial Velocity Orbital Fits\label{tab:orbit}}
\tablehead{
	\colhead{Object} & \colhead{$P$} & \colhead{alias}
   	& \colhead{$k$} & \colhead{$\gamma$} \\
	& (d) & (d) & ({\rm km~s}^{-1}) & ({\rm km~s}^{-1})
}
	\startdata
  J0421+00 & 0.705822 & none     & 86.2^{+ 3.2}_{- 3.3} & -15.8\pm 2.7 \\
  J0750+40 & 1.181426 & none     & 36.2^{+10.5}_{-10.6} & -48.8\pm 8.9 \\
  J0803+41 & 1.344415 & none     & 63.1^{+12.0}_{-17.5} &  57.2\pm 8.9 \\
J0811$-$02 & 1.407864 & 1.32 to 1.47 & 76.5^{+ 7.8}_{-10.1} & 23.9\pm11.2 \\
  J1541+31 & 0.698590 & 1.217964 & 88.8^{+ 5.4}_{- 7.8} &  54.9\pm24.5 \\
  J1600+27 & 1.003368 & 0.962638 & 142.0^{+33.9}_{-52.2}& -293.7\pm52.0 \\
	\enddata
\end{deluxetable}

\begin{deluxetable*}{rccccc} 
\tabletypesize{\footnotesize}
\tablecolumns{6}
\tablewidth{0pt}
\tablecaption{Light Curve Fits for Six Eclipsing Systems\label{tab:lc}}
\tablehead{ 
	\colhead{Object}& \colhead{$i$}& \colhead{$P$}& \colhead{$r_{\rm 1}/a$}& 
	\colhead{$r_{\rm 2}/a$}& \colhead{$L_{\rm 2}/L_{\rm 1}$}\\
	& ($^{\circ}$) & (d) & & &
}
	\startdata 
J0750+40 & 86.8$^{+1.2}_{-3.4}$ & 1.181426(2) & 0.217$^{+0.014}_{-0.049}$ & 0.189$^{+0.028}_{-0.009}$ & 0.096$^{+0.141}_{-0.042}$\\
J0802+43 & 87.2$^{+1.7}_{-4.7}$ & 2.541831(4) & 0.297$^{+0.008}_{-0.038}$ & 0.231$^{+0.028}_{-0.006}$ & \nodata \\
J0803+41 & 83.0$^{+2.5}_{-4.3}$ & 1.344415(4) & 0.269$^{+0.025}_{-0.029}$ & 0.192$^{+0.050}_{-0.016}$ & \nodata \\
J0823+37 & 86.7$^{+1.7}_{-4.6}$ & 0.7339255(7) & 0.275$^{+0.012}_{-0.046}$ & 0.220$^{+0.031}_{-0.009}$ & 0.091$^{+0.125}_{-0.045}$ \\
J0832+13 & 85.4$^{+2.9}_{-3.3}$ & 3.66975(3) & 0.226$^{+0.025}_{-0.017}$ & 0.107$^{+0.018}_{-0.010}$ & \nodata \\
J1600+27 & 86.2$^{+2.8}_{-1.7}$ & 1.003368(2) & 0.108$^{+0.007}_{-0.012}$ & 0.220$^{+0.013}_{-0.016}$ & 1.185$^{+0.101}_{-0.089}$ \\
	\enddata \end{deluxetable*}

\subsection{Spectral Energy Distributions}

\begin{figure} 
  \includegraphics[width=3.5in]{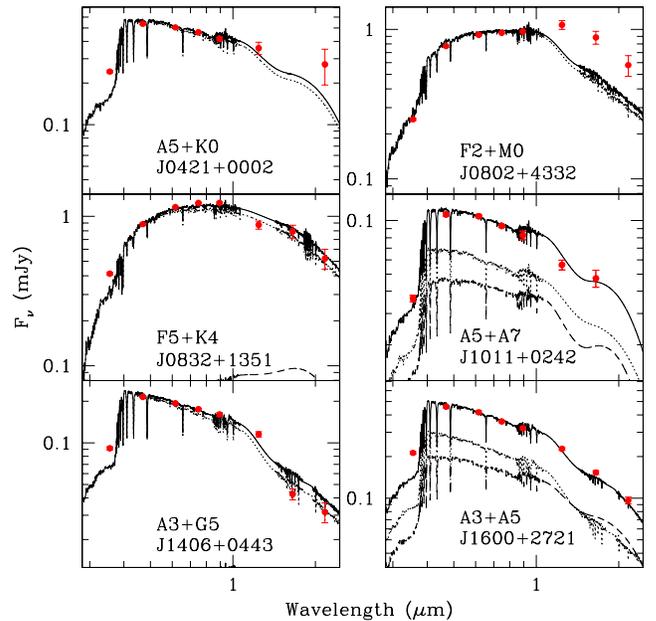} 
	\caption{ \label{fig:sed} Spectral energy distributions for 6 objects with 
infrared photometry in more than one passband.  The SDSS plus 2MASS or UKIDSS 
photometry (red points) are compared against fiducial spectral templates (black 
lines).  The spectral types listed here are representative; the next Figure 
incorporates radial velocity and eclipse information.
	}
\end{figure}

	We search the Two Micron All Sky Survey \citep[2MASS,][]{skrutskie06} and 
the UKIRT Infrared Deep Sky Survey \citep[UKIDSS,][]{lawrence07} and find 
near-infrared photometry in more than one passband for six objects.  Figure 
\ref{fig:sed} shows the spectral energy distributions (SEDs) of these six objects.  
Points with error bars show the observed photometry, whereas the dotted and dashed 
lines show fiducial spectral templates for A0 to M0 type stars \citep{pickles98}.

	Five of these stars, excluding J1011+0242, are binary systems with radial 
velocity variations or photometric eclipses.  Thus we model these SEDs with 
composite A0 to M0 main-sequence star spectral templates, scaling each spectral 
template based on its absolute $V$-band magnitude from Table 2 of \citet{pickles98}.  
The solid lines show the best-fitting composite template for each system.  The SED 
for J0802+4332 is the most unusual in this sample, as it is brightest in the 
$J-$band and has the largest residuals.  J0802+4332 may require a larger and more 
evolved companion to explain its infrared photometry.

	We note that the spectral types shown in Figure \ref{fig:sed} are meant to 
be representative.  The implied mass ratios, based purely on broadband photometry, 
range from $q=0.3$ to 0.9.  Adding the radial velocity and eclipse information 
allows for more accurate constraints.

\section{RESULTS}

	We now combine the data into a joint constraint on the physical nature of 
sdA stars.  Eclipses constrain the ratio of stellar radii and, given the 
Stephan-Boltzmann Law, the ratio of stellar temperatures.  Radial velocities, given 
Kepler's 3rd Law, constrain the ratio of masses in each binary.  Adopting an 
absolute measurement, such as the mass of the primary star, sets the absolute scale 
for the entire system.

	Stellar evolution models provide us with libraries of physically possible
tracks of mass, radius, luminosity, and temperature as a function of a star's age.  
We consider two families of stellar evolution models relevant to these $\log{g}=$ 4
to 6 objects:  Padova main sequence tracks \citep{bressan12} and helium-core WD
tracks \citep{althaus13}.

	Our goal is to identify the combination of primary and secondary stellar 
mass, radius, luminosity, and temperature that, together, best match the 
observations. The observations are orbital period, velocity amplitude, inclination, 
stellar radius-to-semi-major axis ratio, luminosity ratio, and primary temperature.  
We ignore our \logg\ measurements, because surface gravity is the parameter we are 
trying to understand.

	Our approach is to adopt a trial primary mass $M_1$, and then calculate the 
resulting $M_2$, stellar radii, surface gravities, and luminosities required by the 
observations.  Stepping $M_1$ from 0.1 \msun\ to 2.5 \msun\ presents us with the 
plausible range of solutions for the \teff\ $\simeq$ 8000~K primary stars.  The 
correct solution is the one at which the primary and secondary stellar parameters 
self-consistently match evolutionary tracks. We consider the possibility that the 
binaries contain WD+WD, WD+MS, MS+WD, or MS+MS stars.

	Some observational quantities are better constrained than others, and so we 
perform a Monte Carlo calculation to account for the errors.  The result of the 
Monte Carlo calculation is a cloud of possible primary and secondary stellar 
parameters for each system.  For every set of parameters, we record the 
best-matching pair of points in the respective stellar evolutionary tracks.

\begin{figure*} 
\centerline{
 \includegraphics[width=2.7in]{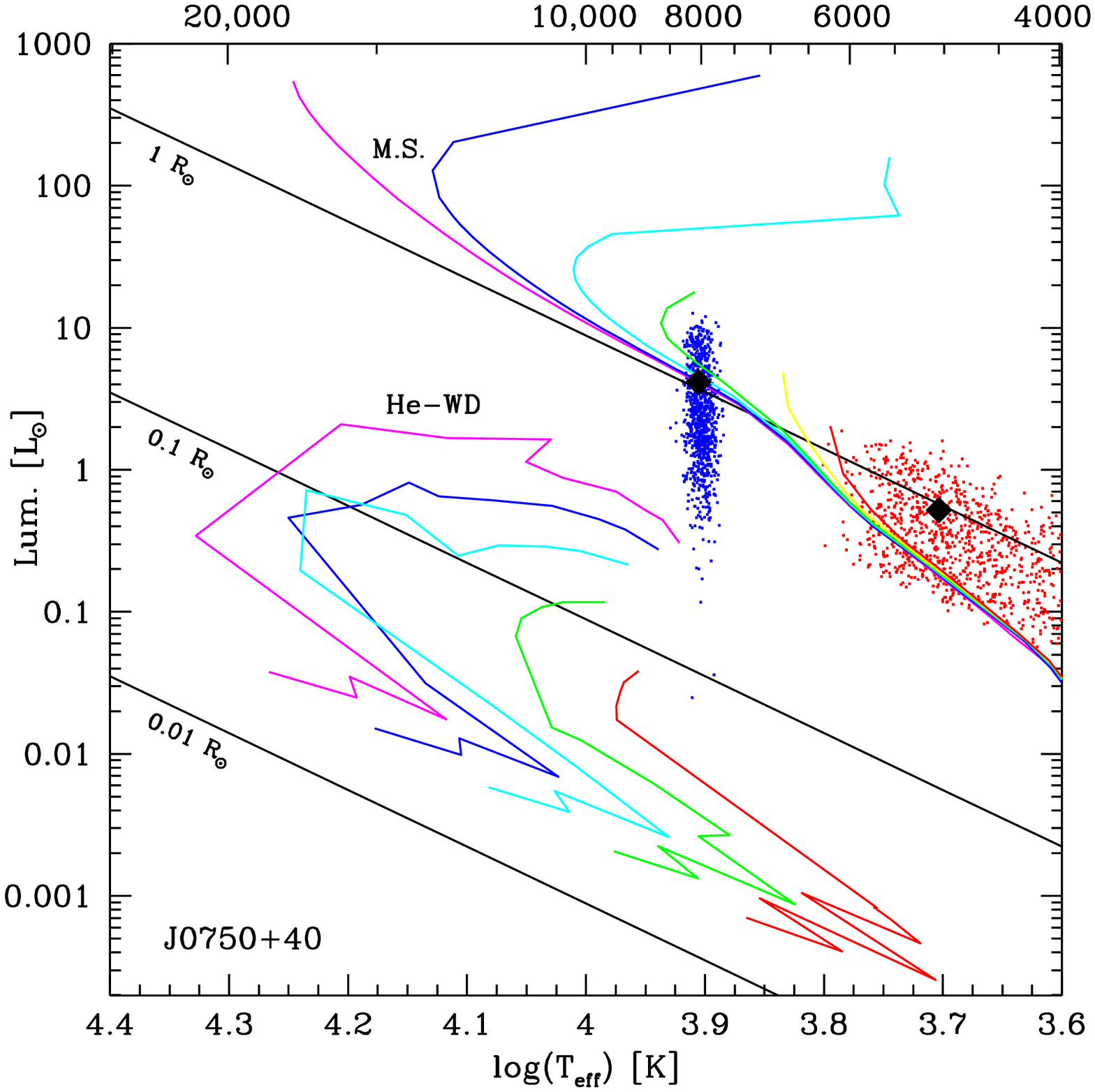} 
 \includegraphics[width=2.7in]{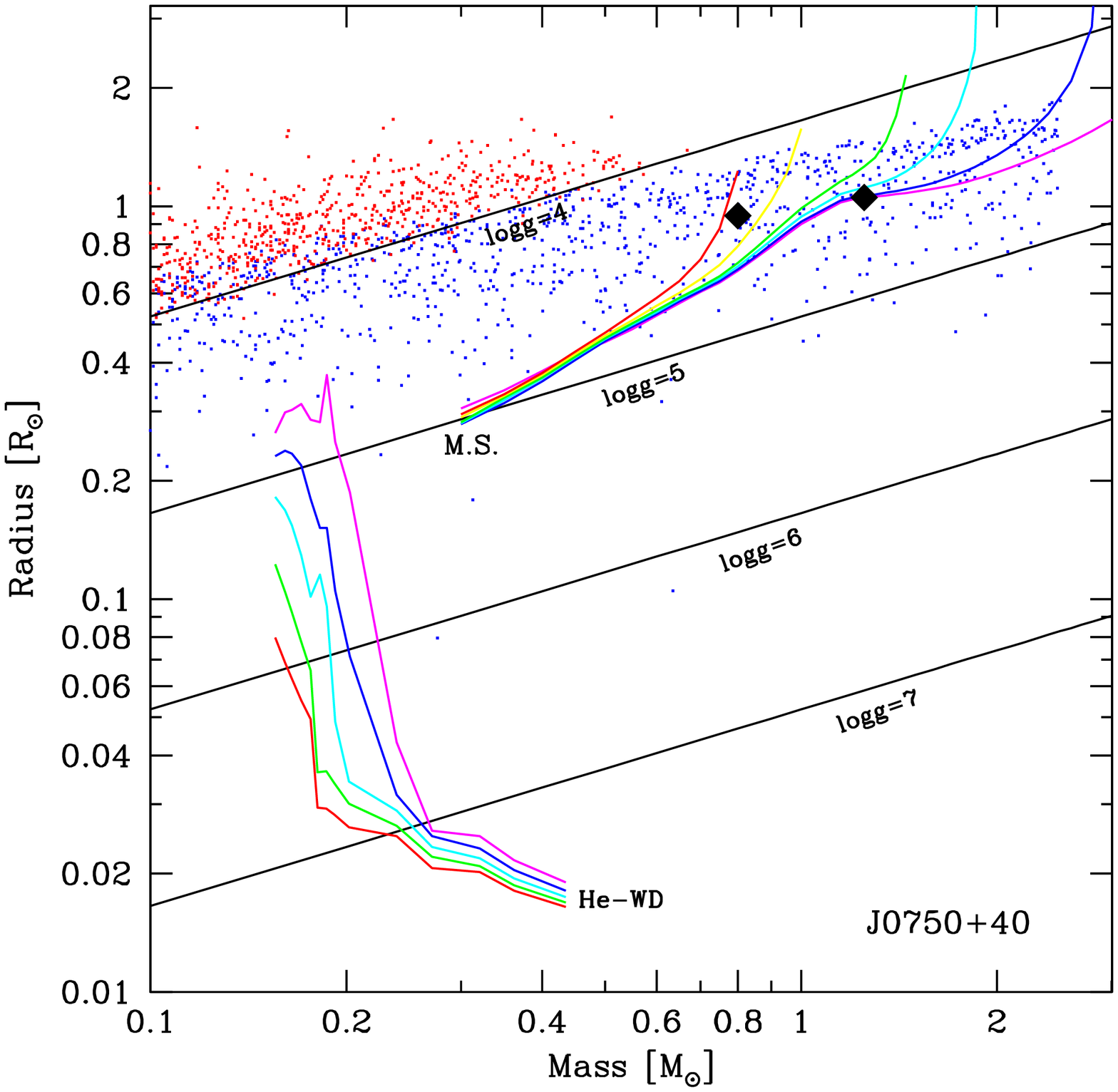}
} \centerline{ 
 \includegraphics[width=2.7in]{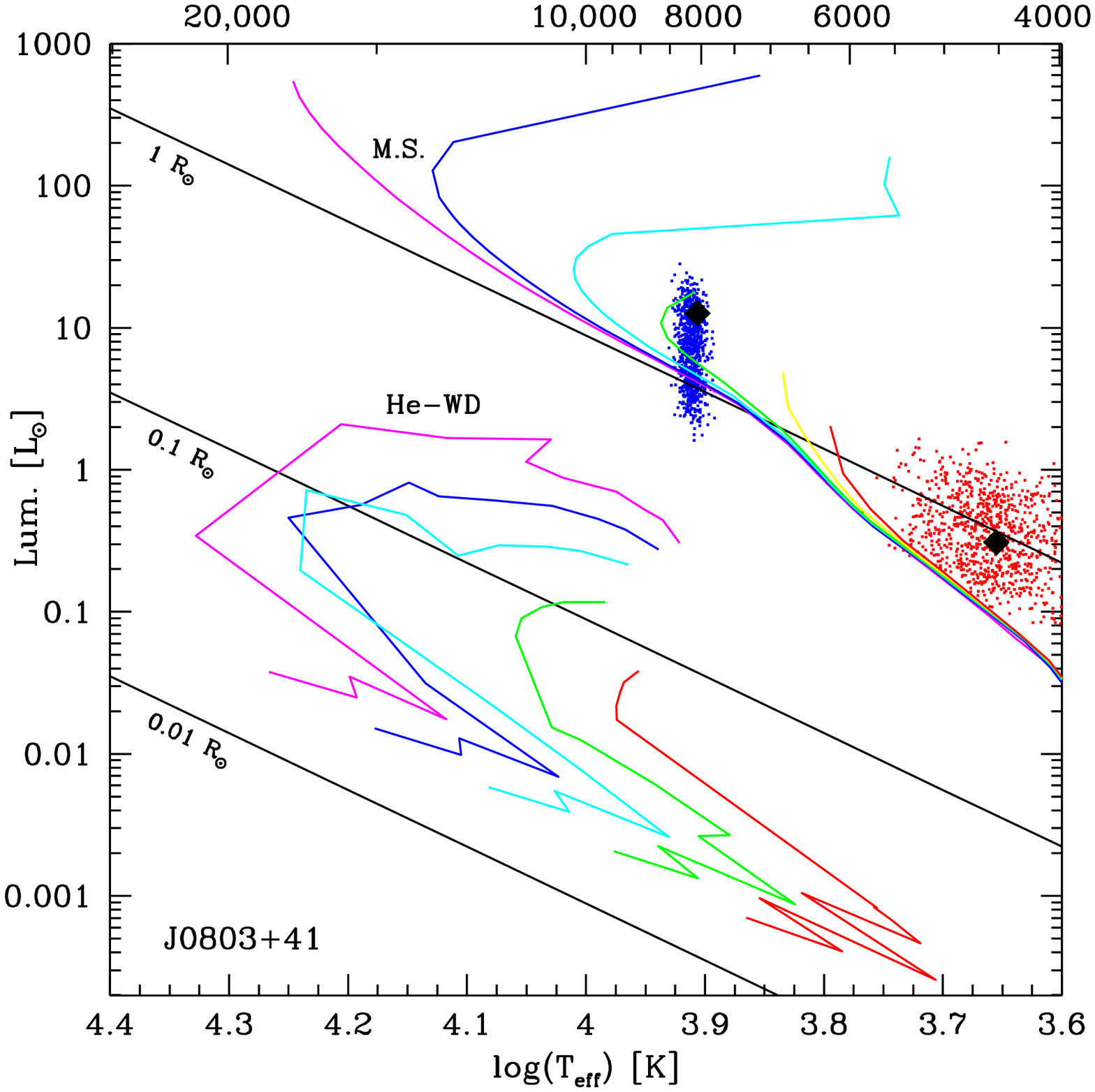} 
 \includegraphics[width=2.7in]{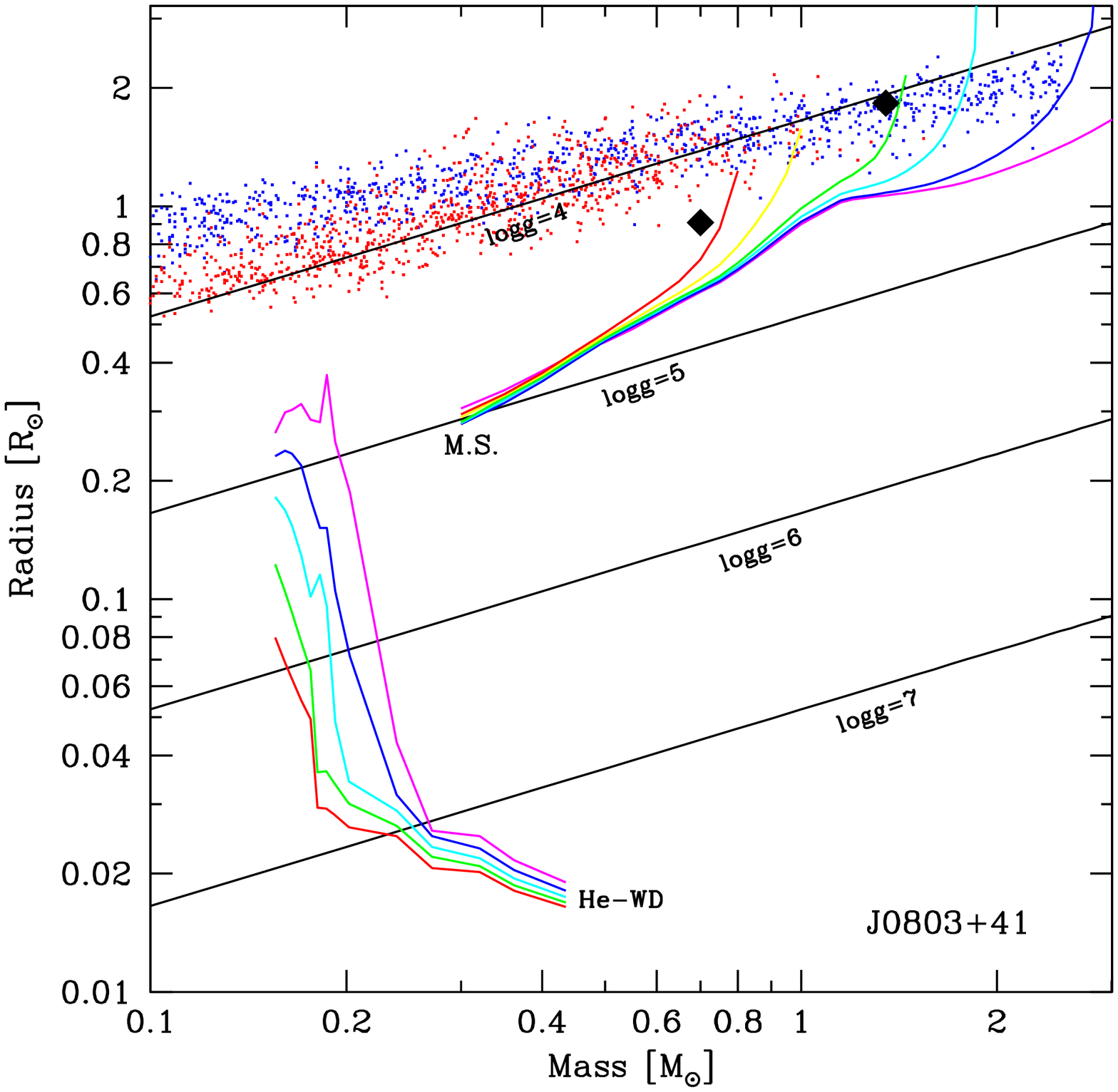}
} \centerline{ 
 \includegraphics[width=2.7in]{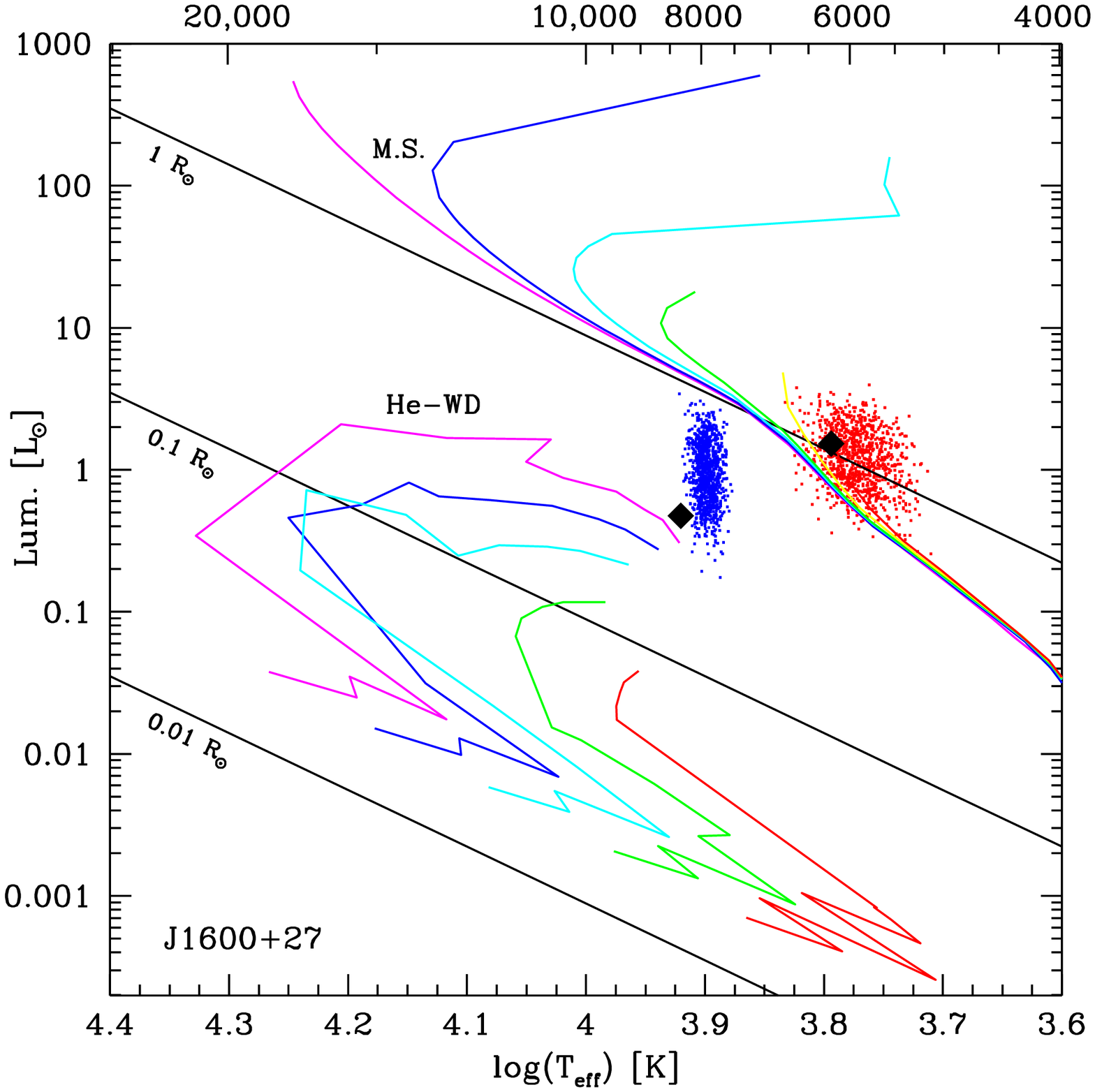} 
 \includegraphics[width=2.7in]{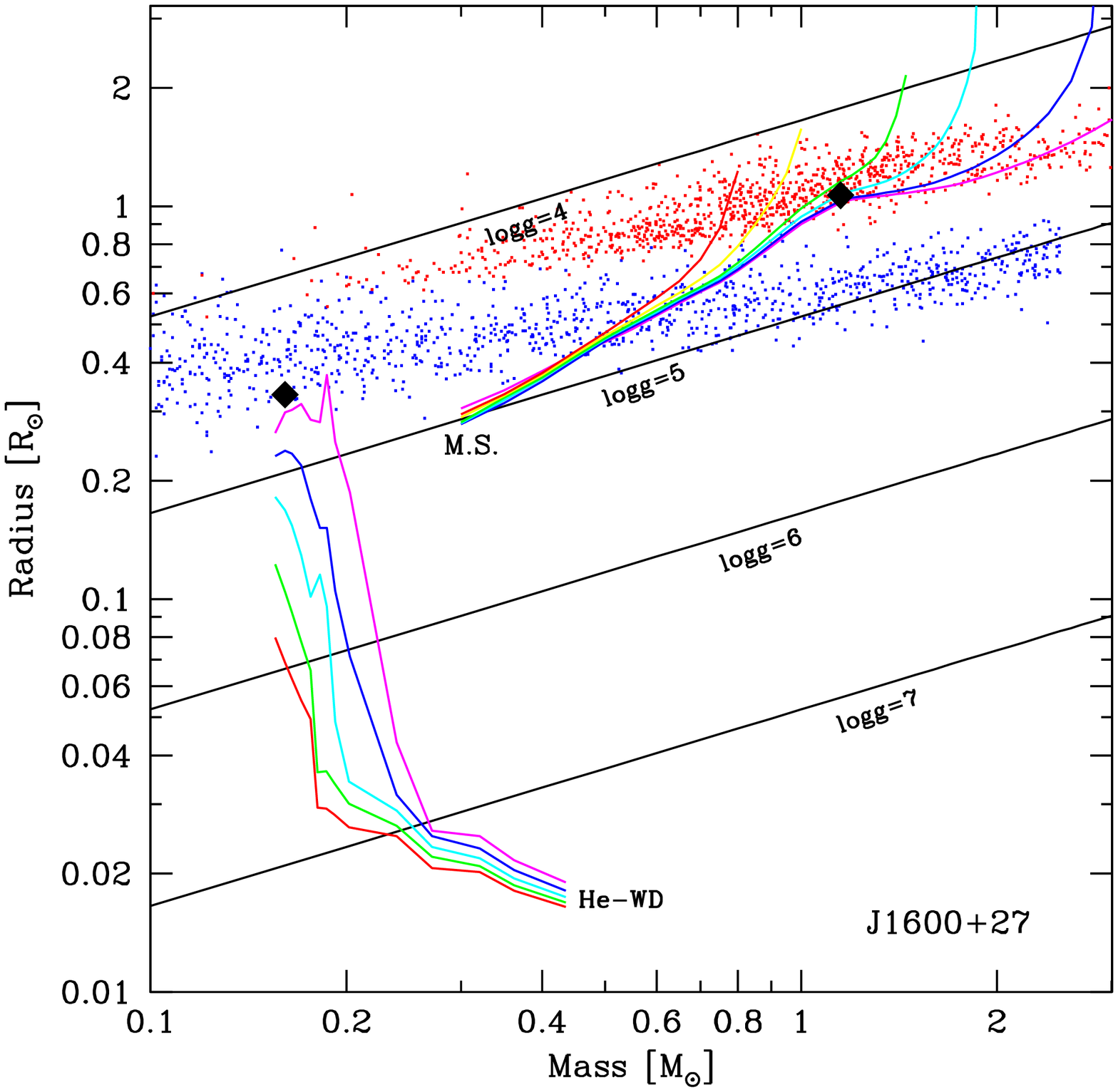}
}
 \caption{ \label{fig:results}
	Temperature-luminosity (left) and mass-radius (right) diagrams showing the 
joint constraints on J0750+40 (top), J0803+41 (middle), and J1600+27 (bottom).  
Possible primary parameters (blue dots) and secondary parameters (red dots) are 
shown for a uniform prior on $M_1$. Colored curves are selected isochrones from 
Padova [M/H]=$-1$ tracks ranging from 0.1 Gyr to 12 Gyr (upper right) and from 
\citet{althaus13} helium-core WD tracks ranging from 0.1 Gyr to 2.4 Gyr (lower 
left).  Black diamonds are the result:  the primary and secondary parameter pair 
that most closely matches the full set of tracks simultaneously in mass, radius, 
temperature, and luminosity.
	}
\end{figure*}

\subsection{Joint Constraints}

	Figure \ref{fig:results} visualizes the results for the three 
best-constrained cases:  J0750+40, J0803+41, and J1600+27.  All three are 8000~K 
objects in eclipsing binaries with radial velocity orbital solutions.  We present 
the results with a pair of plots for each object: temperature-luminosity (left-hand 
panel) and mass-radius (right-hand panel).  All plots are drawn on a log-log scale.  
Thus stellar radius falls along straight lines in the temperature-luminosity plot 
(per Equation 1), and surface gravity falls along straight lines in the mass-radius 
plot (per Equation 2).

	The clouds of blue and red dots in Figure \ref{fig:results} are the results 
of the Monte Carlo calculations, showing the observational constraints on possible 
primary (blue) and secondary (red) parameters assuming a uniform prior on 
$\log{(M_1)}$.  The black diamonds are the result: the parameter pair that 
best-matches with stellar evolutionary tracks simultaneously in mass, radius, 
temperature, and luminosity.

	The colored curves in Figure \ref{fig:results} are selected isochrones from 
the two sets of stellar evolutionary models.  The upper set of isochrones come from 
the Padova [M/H]$=-1$ tracks, stepping from 0.1 Gyr (magenta) to 12 Gyr (red) in 
factors of 3 in age.  The lower set of isochrones come from the \citet{althaus13} WD 
tracks, stepping from 0.15 Gyr (magenta) to 2.4 Gyr (red) in factors of 2 in age.  
Because \citet{althaus13} start their tracks at the moment the WD progenitor 
detaches from the common envelope phase, the WDs begin hot and extended before 
cooling and shrinking with age.  Note that the selected isochrones in Figure 
\ref{fig:results} are for display purposes; we search every time step of every track 
in our comparison.

	We find that the observations of J0750+40 and J0803+41, sdA stars found by 
Kepler et al., are best explained by the metal-poor main sequence tracks:  
$\simeq$1.2~\msun\ stars with $\simeq$0.8~\msun\ companions.  Solutions involving 
WDs yield 25 times larger residuals.  We note that the best-matching surface gravity 
for J0750+40, $\log{g}=4.3$, is in perfect agreement with its SSPP measurement.  If 
we were to use its SSPP effective temperature instead of our pure hydrogen 
model-derived \teff, the best solution to J0750+40 shifts slightly to a 1.15 \msun\
and 0.7 \msun\ pair of metal-poor main sequence stars.

	J1600+27, one of our sdA-like ELM WD candidates, is ambiguous.  Formally, 
the best solution is a 0.16 \msun\ helium-core WD orbiting a $\simeq$0.8 \msun\ main 
sequence companion.  J1600+27's light curve fit yields a companion star that is more 
luminous and larger in radius than the hot ``primary,'' which means the hot primary 
could be a WD.  However, this would indicate that we caught J1600+27 right after 
the WD progenitor detached from the common envelope phase, which seems unlikely. 
Parameter solutions involving metal poor 0.95 \msun\ + 0.95 \msun\ main sequence 
stars are also possible with only 3 times larger residuals, in better agreement with 
J1600+27's SED.  Higher resolution spectroscopy would resolve the issue, as an equal 
mass system would be a double-lined spectroscopic binary.

	Two other eclipsing sdA stars that have infrared excess are J0802+43 and 
J0832+13.  Neither system shows significant radial velocity variability.  Instead, 
we perform our Monte Carlo calculation using our radial velocities as upper limits 
on the velocity amplitudes.  We find that the observations are best explained by 
main sequence tracks:  $\simeq$1.9 \msun\ stars with $\simeq$1.0 \msun\ companions 
if solar metallicity, or $\simeq$1.3 \msun\ stars with $\simeq$0.8 \msun\ companions 
if metal-poor.  The upper limits on velocity push the solutions towards more massive 
stars, and solar metallicity stars are more massive than metal-poor stars of 
identical temperature.  However SSPP reports [Fe/H]$=-0.971\pm0.012$ for J0802+43 
(see Table \ref{tab:teff}) which clearly points to the metal-poor solution.  
Parameter solutions involving WDs have 1000 times larger residuals than main 
sequence star solutions.

	In summary, we consider the joint constraints on five objects with eclipses, 
radial velocity orbits, and/or infrared excess.  The four objects identified as sdA 
stars by Kepler are metal-poor $\simeq$1.2 \msun\ stars with $\simeq$0.8 \msun\ 
companions; the object selected as an ELM WD candidate is either a low mass WD with 
a main sequence companion or an equal-mass main sequence binary.

\subsection{Comparison with Other Results}

	The photometric variability study of \citet{bell17} provides an interesting 
comparison.  They target nine $\sim$8000~K ELM WD candidates suspected to fall in 
the empirical WD instability strip.  Three objects pulsate: two may be ELM WD 
pulsators; one is clearly a high-amplitude $\delta$ Scuti variable.  Two other 
objects are photometric binaries: one has ellipsoidal variations consistent with an 
ELM WD binary; the other has $\simeq$4 hr eclipses and must be a main sequence star.  
They conclude that ELM Survey objects with sdA temperatures are a mix of WDs and 
main sequence stars.

	It is also interesting to check the kinematics of the five eclipsing 
binaries studied here.  An 8000~K metal-poor main sequence star with absolute 
magnitude $M_g\simeq2.5$ is a 1000 times more luminous than an 8000~K ELM WD with 
$M_g\simeq10$ \citep{bressan12, althaus13}.  The objects studied here have a typical 
apparent magnitude of $g=18$.  Thus any metal-poor 1.2 \msun\ star would be about 
13 kpc distant in the halo, while a ELM WD would be about 0.4 kpc distant and in the 
disk.

	Systemic radial velocities for the eclipsing binaries J0832+13 and J1600+27 
are +135 \kms\ and $-294$ \kms, respectively, consistent with halo stars.  Systemic 
radial velocities for the other three eclipsing binaries are around 50 \kms, and 
thus ambiguous.  Proper motions for the five eclipsing binaries are all about 
$3\pm3$ \mas , consistent with zero \citep{altmann17}.  Small proper motions are 
expected for distant stars, however the large inferred $\sim$200 \kms 
tangential velocities are statistically insignificant given the errors.  Thus the 
kinematics are consistent with halo stars, but not a significant constraint.

	Taking this argument in the other direction, an sdA star with a high proper 
motion may be the signature of a nearby ELM WD.  There are 1,364 sdA stars in 
the full \citet{kepler15, kepler16} sample detected in at least 6 epochs in the 
HSOY proper motion catalog; 323 have non-zero proper motions at $>$3-$\sigma$ 
confidence \citep{altmann17}.  If we assume a fiducial absolute magnitude $M_g=2.5$ 
appropriate for a metal-poor 8000~K main sequence star, then 35 (or 3\%) of 
the putative halo sdA stars are unbound to the Milky Way at $>$3-$\sigma$ 
confidence.  We consider unbound tangential velocities implausible given that the 
stars all have modest radial velocities.  It is much more likely that these 35 sdA 
stars are WDs with $M_g\simeq10$, in which case their tangential velocities are 
comparable to their radial velocities.  Including distance errors would reduce the 
number of significant tangential velocity outliers, but, on the basis of proper 
motions, we conclude that a few percent of the sdA objects may be WDs.

\section{DISCUSSION}

	The eclipsing sdA binaries prove one thing:  the sdA population contains 
metal-poor main sequence stars.  ELM WDs that have cooled to A-F type temperatures, 
objects like the eclipsing system NLTT~11748 \citep{steinfadt10, kawka10, kilic10}, 
are presumably mixed into the sdA population.  But what fraction of sdA stars are 
metal-poor main sequence stars, and what fraction are ELM WDs?  The eclipsing 
binaries cannot answer that question, because we do not know if the binaries fairly 
sample the sdA population.  Instead, we turn to theoretical evolutionary tracks and 
the magnitude-limited ELM WD sample.

	Evolutionary tracks provide a precise constraint on the time an ELM WD
spends at sdA-like temperatures.  The accuracy is complicated by
thermonuclear hydrogen shell flashes that alter ELM WD evolutionary times
\citep{driebe98, althaus01}.  The number of shell flashes and the mass threshold at
which they occur depend on assumptions about element diffusion, progenitor
metallicity, and rotation \citep{althaus15, istrate16}.  We consider the models of
\citet{althaus13} and the models of \citet{istrate16} with rotation.  These two sets
of models agree to within 15\% on ELM WD mass and luminosity at a given \teff\ and
\logg, however the time to reach that temperature can in some cases vary by more
than a factor of 2.

	What is true across all models is that a magnitude-limited sample contains 
fewer cool ELM WDs, because cooler WDs are less luminous than hotter WDs.  
Furthermore, a fraction of ELM WDs must merge with their companions before they cool 
to $\sim$8000~K temperatures.  We account for mergers by assuming that ELM WDs have 
formed continuously over the past Gyr with the orbital period distribution derived 
by \citet{brown16b}.  Following the \citet{brown16b} approach to evolving the ELM WD 
population, the ratio of $6500 < T_{\rm eff} < 9000$~K to $10,000< T_{\rm eff} 
< 15,000$~K ELM WDs in a magnitude-limited sample is about 1:2 averaged over the two 
sets of evolutionary models.

	The ELM Survey has now obtained a spectrum for every star in SDSS in the 
magnitude range $15<g<20$ within the ELM Survey color cut.  While some ELM WDs must 
fall outside the color cut, the ELM Survey provides a robust lower limit on the 
absolute number of ELM WDs.  The clean sample of ELM WD binaries contains 36 objects 
with $10,000< T_{\rm eff} < 15,000$~K.  Assuming that the true absolute number of 
ELM WDs in the SDSS footprint is a few times larger than the published number, the 
evolutionary model scaling predicts about 50 ELM WDs with $6500 < T_{\rm eff} 
< 9000$~K in the magnitude range $15<g<20$ in the SDSS footprint.

	By comparison, \citet{kepler15, kepler16} identify about 2600 sdA stars with 
$6500 < T_{\rm eff} < 9000$~K and $5<\log{g}<6.5$ in SDSS.  The sdA spectroscopic 
sample is incomplete.  Thus cool ELM WDs comprise no more than about 2\% of 
the sdA population.  This estimate is consistent with the number of tangential 
velocity outliers in the sdA sample, a completely independent constraint.  We 
conclude that most sdA stars, based on the observations presented here, are likely 
metal-poor A- and F-type main sequence stars in the halo.

\subsection{Implications}

	The physical nature of sdA stars has important implications for their binary 
properties.  The minimum orbital period for a detached binary is set by the Roche 
lobe radius \citep{eggleton83}.  If we adopt $q=0.67$ seen in the eclipsing 
binaries, and assume that all sdA stars are $\simeq$1.2~\msun\ metal-poor stars, 
then the Roche lobe criterion demands that no sdA binary has $P<9$ hr.

	Indeed, no $P<9$ hr system has yet been found among the sdA stars.  Finding 
a $P<9$ hr system would be evidence that the binary contains a WD.  If sdA stars 
were solar metallicity stars, on the other hand, the period minimum is about 24 hr.  
This is already ruled out by J0421+00, J0823+37, and possibly J1541+31.  

	Two-thirds of ELM WDs are observed in detached binaries with orbital periods 
$P<9$ hr \citep{brown16b}, binaries that cannot fit sdA stars.  The implication is 
clear:  the Roche lobe criterion sets a minimum binary orbital period for sdA stars 
50 times larger than that of ELM WDs.

	The physical nature of sdA stars also has implications for studies of 
pulsating ELM WDs.  First discovered by \citet{hermes12b, hermes13a}, the first 
three ELM WD pulsators are hotter than 9000~K and are in binary systems with $P=4.1$ 
hr to 14.6 hr.  The pulsating ELM WD companion to the millisecond pulsar PSR 
J1738+0333 has similar properties \citep{kilic15}.  The temperatures and orbital 
periods of these pulsators are consistent with being ELM WDs.

	However, the fourth and fifth members of this class of pulsators are cooler 
than 9000~K and do not show any significant radial velocity variations 
\citep{hermes13b}. \citet{corti16} and \citet{bell17} present five additional 
pulsators with temperatures below 8000~K, and with no evidence of binarity.
	Because pulsations are common in A-type stars, it is possible that these 
relatively cool pulsators are SX~Phe variables, and not pulsating ELM WDs.  Hence, 
there may be only four likely pulsating ELM WDs:  the three pulsators presented in 
\citet{hermes13a} and the WD companion to PSR J1738+0333 \citep{kilic15}.  
Measuring the rate of pulsation period change could clarify their stellar nature, 
because different classes of pulsators have different rates of period change 
\citep{calcaferro17}.

\section{Summary}

	In this paper we investigate the physical nature of sdA stars and their 
possible link to ELM WDs.  The distribution of colors and reduced proper motions 
indicate that sdA stars are cooler and more luminous, and thus larger in radius, 
than published ELM WDs.  We perform a detailed study of sdA stars in eclipsing 
binaries with infrared excess and/or radial velocity orbital solutions.  The joint 
observational constraints are best explained with binaries containing metal-poor 
$\simeq$1.2 \msun\ main sequence stars with $\simeq$0.8 \msun\ companions, not with 
ELM WDs.

	The source of confusion comes from fitting pure hydrogen models to sdA 
spectra.  Metal line blanketing is important below 9000~K, and the Balmer lines 
become insensitive to temperature \citep{stromgren69}.  We demonstrate that pure 
hydrogen model fits to synthetic A-F type spectra yield systematically wrong surface 
gravities by $\sim$1 dex.  Empirically, surface gravities derived by SSPP for a set 
of sdA stars differ by 1.8 dex compared to their pure hydrogen model derived values. 
Thus sdA stars fit with pure hydrogen models appear to be WD imposters.

	While it is true that ELM WDs must exist at sdA-like temperatures, they are
intrinsically faint and sufficiently rare that we predict about 50 cool ELM
WDs with $15<g<20$ mag in the SDSS survey.  Thus ELM WDs comprise of order
1\% of the observed sdA population.  This conclusion is supported by the
small number of sdA tangential velocity outliers, and the absence of $P<9$ hr
periods in observed sdA binaries.  The majority of sdA stars are likely metal-poor
A--F type stars in the halo.


\acknowledgements

	We thank S.\ O.\ Kepler for providing us with the list of sdA eclipsing 
binary candidates.  We thank B. Kunk, E.\ Martin, and A.\ Milone for their 
assistance with observations obtained at the MMT Observatory, P.\ Canton for his 
assistance with observations obtained at Kitt Peak National Observatory, and P.\ 
Berlind and M.\ Calkins for their assistance with observations obtained at the Fred 
Lawrence Whipple Observatory. MK and AG gratefully acknowledge the support of the 
NSF and NASA under grants AST-1312678 and NNX14AF65G. This project makes use of data 
products from the Sloan Digital Sky Survey, which is managed by the Astrophysical 
Research Consortium for the Participating Institutions.  This project makes use of 
data products from the Catalina Sky Surveys.  The CSS survey is funded by the 
National Aeronautics and Space Administration under Grant No. NNG05GF22G issued 
through the Science Mission Directorate Near-Earth Objects Observations Program.  
The CRTS survey is supported by the U.S.~National Science Foundation under grants 
AST-0909182 and AST-1313422. This research makes use the SAO/NASA Astrophysics Data 
System Bibliographic Service.  This work was supported in part by the Smithsonian 
Institution.

\facilities{MMT (Blue Channel Spectrograph), Mayall (KOSMOS), FLWO:1.5m (FAST)}

\appendix \section{DATA TABLE}

	Table \ref{tab:dat} presents the radial velocity measurements for the sample 
of 11 sdA stars suspected of being eclipsing binaries and 11 ELM WD candidates that 
have sdA-like temperatures.  Table \ref{tab:dat} columns include object name, 
heliocentric Julian date (based on UTC), heliocentric radial velocity (uncorrected 
for the WD gravitational redshift), and velocity error.

\begin{deluxetable}{clC}        
\tablecolumns{3}
\tablewidth{0pt}
\tablecaption{Radial Velocity Data\label{tab:dat}}
\tablehead{
	\colhead{Object}& \colhead{HJD} & \colhead{$v_{\rm helio}$}\\
			& +2450000 d	& ({\rm km~s}^{-1}) }
	\startdata
  J0747+45 & 7336.949516 &   64.11 \pm  5.20 \\
   \nodata & 7337.022786 &   60.35 \pm  5.05 \\
   \nodata & 7427.696022 &   84.46 \pm  8.43 \\
   \nodata & 7427.859036 &   87.48 \pm  8.18 \\
   \nodata & 7428.695398 &   99.08 \pm  7.05 \\
   \nodata & 7430.691555 &   73.15 \pm  7.84 \\
  J0750+40 & 7336.962809 &  -20.97 \pm  4.66 \\
   \nodata & 7337.017686 &  -37.19 \pm  7.11 \\
   \nodata & 7427.717586 &  -12.25 \pm  6.72 \\
   \nodata & 7427.852464 &  -31.56 \pm  7.02 \\
	\enddata \tablecomments{This table is available in its entirety in
machine-readable and Virtual Observatory forms in the online journal. A portion is
shown here for guidance regarding its form and content.}
\end{deluxetable}

\clearpage


\begin{thebibliography}{}
\expandafter\ifx\csname natexlab\endcsname\relax\def\natexlab#1{#1}\fi
\providecommand{\url}[1]{\href{#1}{#1}}

\bibitem[{{Abt} {et~al.}(2002){Abt}, {Levato}, \& {Grosso}}]{abt02}
{Abt}, H.~A., {Levato}, H., \& {Grosso}, M. 2002, \apj, 573, 359

\bibitem[{{Abt} \& {Morrell}(1995)}]{abt95}
{Abt}, H.~A., \& {Morrell}, N.~I. 1995, \apjs, 99, 135

\bibitem[{{Allende Prieto} {et~al.}(2014){Allende Prieto},
  {Fern{\'a}ndez-Alvar}, {Schlesinger}, {et~al.}}]{allende14}
{Allende Prieto}, C., {Fern{\'a}ndez-Alvar}, E., {Schlesinger}, K.~J., {et~al.}
  2014, \aap, 568, A7

\bibitem[{{Allende Prieto} {et~al.}(2008){Allende Prieto}, {Sivarani}, {Beers},
  {et~al.}}]{allende08}
{Allende Prieto}, C., {Sivarani}, T., {Beers}, T.~C., {et~al.} 2008, \aj, 136,
  2070

\bibitem[{{Althaus} {et~al.}(2015){Althaus}, {Camisassa}, {Miller Bertolami},
  {C{\'o}rsico}, \& {Garc{\'{\i}}a-Berro}}]{althaus15}
{Althaus}, L.~G., {Camisassa}, M.~E., {Miller Bertolami}, M.~M., {C{\'o}rsico},
  A.~H., \& {Garc{\'{\i}}a-Berro}, E. 2015, \aap, 576, A9

\bibitem[{{Althaus} {et~al.}(2013){Althaus}, {Miller Bertolami}, \&
  {C{\'o}rsico}}]{althaus13}
{Althaus}, L.~G., {Miller Bertolami}, M.~M., \& {C{\'o}rsico}, A.~H. 2013,
  \aap, 557, A19

\bibitem[{{Althaus} {et~al.}(2001){Althaus}, {Serenelli}, \&
  {Benvenuto}}]{althaus01}
{Althaus}, L.~G., {Serenelli}, A.~M., \& {Benvenuto}, O.~G. 2001, \mnras, 324,
  617

\bibitem[{{Altmann} {et~al.}(2017){Altmann}, {Roeser}, {Demleitner}, {Bastian},
  \& {Schilbach}}]{altmann17}
{Altmann}, M., {Roeser}, S., {Demleitner}, M., {Bastian}, U., \& {Schilbach},
  E. 2017, \aap, accepted

\bibitem[{{Bell} {et~al.}(2017){Bell}, {Gianninas}, {Hermes},
  {et~al.}}]{bell17}
{Bell}, K.~J., {Gianninas}, A., {Hermes}, J.~J., {et~al.} 2017, \apj, submitted

\bibitem[{{Breger}(2000)}]{breger00}
{Breger}, M. 2000, in ASP Conf.\ Ser.\, Vol. 210, Delta Scuti and Related
  Stars, ed. M.~{Breger} \& M.~{Montgomery} (San Francisco: ASP), 3

\bibitem[{{Bressan} {et~al.}(2012){Bressan}, {Marigo}, {Girardi},
  {et~al.}}]{bressan12}
{Bressan}, A., {Marigo}, P., {Girardi}, L., {et~al.} 2012, \mnras, 427, 127

\bibitem[{{Brown} {et~al.}(2016{\natexlab{a}}){Brown}, {Gianninas}, {Kilic},
  {Kenyon}, \& {Allende Prieto}}]{brown16a}
{Brown}, W.~R., {Gianninas}, A., {Kilic}, M., {Kenyon}, S.~J., \& {Allende
  Prieto}, C. 2016{\natexlab{a}}, \apj, 818, 155

\bibitem[{{Brown} {et~al.}(2013){Brown}, {Kilic}, {Allende Prieto},
  {Gianninas}, \& {Kenyon}}]{brown13a}
{Brown}, W.~R., {Kilic}, M., {Allende Prieto}, C., {Gianninas}, A., \&
  {Kenyon}, S.~J. 2013, \apj, 769, 66

\bibitem[{{Brown} {et~al.}(2010){Brown}, {Kilic}, {Allende Prieto}, \&
  {Kenyon}}]{brown10c}
{Brown}, W.~R., {Kilic}, M., {Allende Prieto}, C., \& {Kenyon}, S.~J. 2010,
  \apj, 723, 1072

\bibitem[{{Brown} {et~al.}(2012){Brown}, {Kilic}, {Allende Prieto}, \&
  {Kenyon}}]{brown12a}
---. 2012, \apj, 744, 142

\bibitem[{{Brown} {et~al.}(2011){Brown}, {Kilic}, {Hermes},
  {et~al.}}]{brown11b}
{Brown}, W.~R., {Kilic}, M., {Hermes}, J.~J., {et~al.} 2011, \apjl, 737, L23

\bibitem[{{Brown} {et~al.}(2016{\natexlab{b}}){Brown}, {Kilic}, {Kenyon}, \&
  {Gianninas}}]{brown16b}
{Brown}, W.~R., {Kilic}, M., {Kenyon}, S.~J., \& {Gianninas}, A.
  2016{\natexlab{b}}, \apj, 824, 46

\bibitem[{{Calcaferro} {et~al.}(2017){Calcaferro}, {C{\'o}rsico}, \&
  {Althaus}}]{calcaferro17}
{Calcaferro}, L.~M., {C{\'o}rsico}, A.~H., \& {Althaus}, L.~G. 2017, \aap,
  accepted

\bibitem[{{Carter} {et~al.}(2011){Carter}, {Rappaport}, \&
  {Fabrycky}}]{carter11}
{Carter}, J.~A., {Rappaport}, S., \& {Fabrycky}, D. 2011, \apj, 728, 139

\bibitem[{{Castelli} {et~al.}(1997){Castelli}, {Gratton}, \&
  {Kurucz}}]{castelli97}
{Castelli}, F., {Gratton}, R.~G., \& {Kurucz}, R.~L. 1997, \aap, 318, 841

\bibitem[{{Castelli} \& {Kurucz}(2004)}]{castelli04}
{Castelli}, F., \& {Kurucz}, R.~L. 2004, arXiv:astro-ph/0405087

\bibitem[{{Claret} \& {Bloemen}(2011)}]{claret11}
{Claret}, A., \& {Bloemen}, S. 2011, \aap, 529, A75

\bibitem[{{Corti} {et~al.}(2016){Corti}, {Kanaan}, {C{\'o}rsico},
  {et~al.}}]{corti16}
{Corti}, M.~A., {Kanaan}, A., {C{\'o}rsico}, A.~H., {et~al.} 2016, \aap, 587,
  L5

\bibitem[{{Drake} {et~al.}(2009){Drake}, {Djorgovski}, {Mahabal},
  {et~al.}}]{drake09}
{Drake}, A.~J., {Djorgovski}, S.~G., {Mahabal}, A., {et~al.} 2009, \apj, 696,
  870

\bibitem[{{Driebe} {et~al.}(1998){Driebe}, {Schoenberner}, {Bloecker}, \&
  {Herwig}}]{driebe98}
{Driebe}, T., {Schoenberner}, D., {Bloecker}, T., \& {Herwig}, F. 1998, \aap,
  339, 123

\bibitem[{{Eggleton}(1983)}]{eggleton83}
{Eggleton}, P.~P. 1983, \apj, 268, 368

\bibitem[{{Fabricant} {et~al.}(1998){Fabricant}, {Cheimets}, {Caldwell}, \&
  {Geary}}]{fabricant98}
{Fabricant}, D., {Cheimets}, P., {Caldwell}, N., \& {Geary}, J. 1998, \pasp,
  110, 79

\bibitem[{{Gentile Fusillo} {et~al.}(2015){Gentile Fusillo}, {G{\"a}nsicke}, \&
  {Greiss}}]{fusillo15}
{Gentile Fusillo}, N.~P., {G{\"a}nsicke}, B.~T., \& {Greiss}, S. 2015, \mnras,
  448, 2260

\bibitem[{{Gianninas} {et~al.}(2011){Gianninas}, {Bergeron}, \&
  {Ruiz}}]{gianninas11}
{Gianninas}, A., {Bergeron}, P., \& {Ruiz}, M.~T. 2011, \apj, 743, 138

\bibitem[{{Gianninas} {et~al.}(2004){Gianninas}, {Dufour}, \&
  {Bergeron}}]{gianninas04}
{Gianninas}, A., {Dufour}, P., \& {Bergeron}, P. 2004, \apjl, 617, L57

\bibitem[{{Gianninas} {et~al.}(2014{\natexlab{a}}){Gianninas}, {Dufour},
  {Kilic}, {et~al.}}]{gianninas14b}
{Gianninas}, A., {Dufour}, P., {Kilic}, M., {et~al.} 2014{\natexlab{a}}, \apj,
  794, 35

\bibitem[{{Gianninas} {et~al.}(2014{\natexlab{b}}){Gianninas}, {Hermes},
  {Brown}, {et~al.}}]{gianninas14}
{Gianninas}, A., {Hermes}, J.~J., {Brown}, W.~R., {et~al.} 2014{\natexlab{b}},
  \apj, 781, 104

\bibitem[{{Gianninas} {et~al.}(2015){Gianninas}, {Kilic}, {Brown}, {Canton}, \&
  {Kenyon}}]{gianninas15}
{Gianninas}, A., {Kilic}, M., {Brown}, W.~R., {Canton}, P., \& {Kenyon}, S.~J.
  2015, \apj, 812, 167

\bibitem[{{Hermes} {et~al.}(2014){Hermes}, {Brown}, {Kilic},
  {et~al.}}]{hermes14}
{Hermes}, J.~J., {Brown}, W.~R., {Kilic}, M., {et~al.} 2014, \apj, 792, 39

\bibitem[{{Hermes} {et~al.}(2017){Hermes}, {Gaensicke}, \& {Breedt}}]{hermes17}
{Hermes}, J.~J., {Gaensicke}, B.~T., \& {Breedt}, E. 2017, in ASP Conf.\ Ser.:
  20th European Workshop on White Dwarfs, ed. P.-E. {Tremblay}, B.~{Gaensicke},
  \& T.~{Marsh} (San Francisco: ASP), in press

\bibitem[{{Hermes} {et~al.}(2012{\natexlab{a}}){Hermes}, {Kilic}, {Brown},
  {et~al.}}]{hermes12c}
{Hermes}, J.~J., {Kilic}, M., {Brown}, W.~R., {et~al.} 2012{\natexlab{a}},
  \apjl, 757, L21

\bibitem[{{Hermes} {et~al.}(2013{\natexlab{a}}){Hermes}, {Montgomery},
  {Gianninas}, {et~al.}}]{hermes13b}
{Hermes}, J.~J., {Montgomery}, M.~H., {Gianninas}, A., {et~al.}
  2013{\natexlab{a}}, \mnras, 436, 3573

\bibitem[{{Hermes} {et~al.}(2012{\natexlab{b}}){Hermes}, {Montgomery},
  {Winget}, {et~al.}}]{hermes12b}
{Hermes}, J.~J., {Montgomery}, M.~H., {Winget}, D.~E., {et~al.}
  2012{\natexlab{b}}, \apjl, 750, L28

\bibitem[{{Hermes} {et~al.}(2013{\natexlab{b}}){Hermes}, {Montgomery},
  {Winget}, {et~al.}}]{hermes13a}
---. 2013{\natexlab{b}}, \apj, 765, 102

\bibitem[{{Iben}(1990)}]{iben90}
{Iben}, Jr., I. 1990, \apj, 353, 215

\bibitem[{{Istrate} {et~al.}(2016){Istrate}, {Marchant}, {Tauris},
  {et~al.}}]{istrate16}
{Istrate}, A.~G., {Marchant}, P., {Tauris}, T.~M., {et~al.} 2016, \aap, 595,
  A35

\bibitem[{{Kawka} {et~al.}(2010){Kawka}, {Vennes}, \& {Vaccaro}}]{kawka10}
{Kawka}, A., {Vennes}, S., \& {Vaccaro}, T.~R. 2010, \aap, 516, L7

\bibitem[{{Kenyon} \& {Garcia}(1986)}]{kenyon86}
{Kenyon}, S.~J., \& {Garcia}, M.~R. 1986, \aj, 91, 125

\bibitem[{{Kepler} {et~al.}(2015){Kepler}, {Pelisoli}, {Koester},
  {et~al.}}]{kepler15}
{Kepler}, S.~O., {Pelisoli}, I., {Koester}, D., {et~al.} 2015, \mnras, 446,
  4078

\bibitem[{{Kepler} {et~al.}(2016){Kepler}, {Pelisoli}, {Koester},
  {et~al.}}]{kepler16}
---. 2016, \mnras, 455, 3413

\bibitem[{{Kilic} {et~al.}(2010){Kilic}, {Brown}, {Allende Prieto}, {Kenyon},
  \& {Panei}}]{kilic10}
{Kilic}, M., {Brown}, W.~R., {Allende Prieto}, C., {Kenyon}, S.~J., \& {Panei},
  J.~A. 2010, \apj, 716, 122

\bibitem[{{Kilic} {et~al.}(2011){Kilic}, {Brown}, {Allende Prieto},
  {et~al.}}]{kilic11a}
{Kilic}, M., {Brown}, W.~R., {Allende Prieto}, C., {et~al.} 2011, \apj, 727, 3

\bibitem[{{Kilic} {et~al.}(2012){Kilic}, {Brown}, {Allende Prieto},
  {et~al.}}]{kilic12a}
---. 2012, \apj, 751, 141

\bibitem[{{Kilic} {et~al.}(2014){Kilic}, {Brown}, {Gianninas},
  {et~al.}}]{kilic14}
{Kilic}, M., {Brown}, W.~R., {Gianninas}, A., {et~al.} 2014, \mnras, 444, L1

\bibitem[{{Kilic} {et~al.}(2015){Kilic}, {Hermes}, {Gianninas}, \&
  {Brown}}]{kilic15}
{Kilic}, M., {Hermes}, J.~J., {Gianninas}, A., \& {Brown}, W.~R. 2015, \mnras,
  446, L26

\bibitem[{{Kilic} {et~al.}(2008){Kilic}, {Thorstensen}, \& {Koester}}]{kilic08}
{Kilic}, M., {Thorstensen}, J.~R., \& {Koester}, D. 2008, \apjl, 689, L45

\bibitem[{{Kilic} {et~al.}(2006)}]{kilic06}
{Kilic}, M., {et~al.} 2006, \aj, 131, 582

\bibitem[{{Kurtz} \& {Mink}(1998)}]{kurtz98}
{Kurtz}, M.~J., \& {Mink}, D.~J. 1998, \pasp, 110, 934

\bibitem[{{Lawrence} {et~al.}(2007){Lawrence}, {Warren}, {Almaini},
  {et~al.}}]{lawrence07}
{Lawrence}, A., {Warren}, S.~J., {Almaini}, O., {et~al.} 2007, \mnras, 379,
  1599

\bibitem[{{Lee} {et~al.}(2008){Lee}, {Beers}, {Sivarani}, {et~al.}}]{lee08}
{Lee}, Y.~S., {Beers}, T.~C., {Sivarani}, T., {et~al.} 2008, \aj, 136, 2022

\bibitem[{{Marsh} {et~al.}(1995){Marsh}, {Dhillon}, \& {Duck}}]{marsh95}
{Marsh}, T.~R., {Dhillon}, V.~S., \& {Duck}, S.~R. 1995, \mnras, 275, 828

\bibitem[{{Martini} {et~al.}(2014){Martini}, {Elias}, {Points},
  {et~al.}}]{martini14}
{Martini}, P., {Elias}, J., {Points}, S., {et~al.} 2014, in Proc. SPIE 9147,
  Ground-based and Airborne Instrumentation for Astronomy V, ed. {Ramsay, S.~K
  and McLean, I.~S. and Takami, H.} (Montreal, Canada: SPIE), 91470Z

\bibitem[{{Maxted} {et~al.}(2014){Maxted}, {Bloemen}, {Heber},
  {et~al.}}]{maxted14}
{Maxted}, P.~F.~L., {Bloemen}, S., {Heber}, U., {et~al.} 2014, \mnras, 437,
  1681

\bibitem[{{Maxted} {et~al.}(2013){Maxted}, {Serenelli}, {Miglio},
  {et~al.}}]{maxted13}
{Maxted}, P.~F.~L., {Serenelli}, A.~M., {Miglio}, A., {et~al.} 2013, \nat, 498,
  463

\bibitem[{{North} \& {Zahn}(2004)}]{north04}
{North}, P., \& {Zahn}, J.-P. 2004, \nar, 48, 741

\bibitem[{{Payne}(1925)}]{payne25}
{Payne}, C.~H. 1925, PhD thesis, Harvard University

\bibitem[{{Pelisoli} {et~al.}(2017){Pelisoli}, {Kepler}, {Koester}, \&
  {Romero}}]{pelisoli17}
{Pelisoli}, I., {Kepler}, S.~O., {Koester}, D., \& {Romero}, A.~D. 2017, in ASP
  Conf.\ Ser.: 20th European Workshop on White Dwarfs, ed. P.-E. {Tremblay},
  B.~{Gaensicke}, \& T.~{Marsh} (San Francisco: ASP), in press

\bibitem[{{Pickles}(1998)}]{pickles98}
{Pickles}, A.~J. 1998, \pasp, 110, 863

\bibitem[{{Press} {et~al.}(1992){Press}, {Teukolsky}, {Vetterling}, \&
  {Flannery}}]{press92}
{Press}, W.~H., {Teukolsky}, S.~A., {Vetterling}, W.~T., \& {Flannery}, B.~P.
  1992, {Numerical recipes in C. The art of scientific computing} (Cambridge:
  University Press, 2nd ed.)

\bibitem[{{Rappaport} {et~al.}(2015){Rappaport}, {Nelson}, {Levine},
  {et~al.}}]{rappaport15}
{Rappaport}, S., {Nelson}, L., {Levine}, A., {et~al.} 2015, \apj, 803, 82

\bibitem[{{Schmidt} {et~al.}(1989){Schmidt}, {Weymann}, \& {Foltz}}]{schmidt89}
{Schmidt}, G.~D., {Weymann}, R.~J., \& {Foltz}, C.~B. 1989, \pasp, 101, 713

\bibitem[{{Skrutskie} {et~al.}(2006){Skrutskie}, {Cutri}, {Stiening},
  {et~al.}}]{skrutskie06}
{Skrutskie}, M.~F., {Cutri}, R.~M., {Stiening}, R., {et~al.} 2006, \aj, 131,
  1163

\bibitem[{{Smolinski} {et~al.}(2011){Smolinski}, {Lee}, {Beers},
  {et~al.}}]{smolinski11}
{Smolinski}, J.~P., {Lee}, Y.~S., {Beers}, T.~C., {et~al.} 2011, \aj, 141, 89

\bibitem[{{Southworth} {et~al.}(2004){Southworth}, {Maxted}, \&
  {Smalley}}]{southworth04}
{Southworth}, J., {Maxted}, P.~F.~L., \& {Smalley}, B. 2004, \mnras, 351, 1277

\bibitem[{{Steinfadt} {et~al.}(2010){Steinfadt}, {Kaplan}, {Shporer},
  {Bildsten}, \& {Howell}}]{steinfadt10}
{Steinfadt}, J.~D.~R., {Kaplan}, D.~L., {Shporer}, A., {Bildsten}, L., \&
  {Howell}, S.~B. 2010, \apjl, 716, L146

\bibitem[{{Strom}(1969)}]{strom69}
{Strom}, S.~E. 1969, in Theory and Observation of Normal Stellar Atmospheres,
  ed. O.~{Gingerich} (Cambridge: MIT), 99

\bibitem[{{Str{\"o}mgren}(1969)}]{stromgren69}
{Str{\"o}mgren}, B. 1969, in Theory and Observation of Normal Stellar
  Atmospheres, ed. O.~{Gingerich} (Cambridge: MIT), 337

\bibitem[{{Str{\"o}mgren} {et~al.}(1944){Str{\"o}mgren}, {Gyldenkaerne},
  {Rudkjobing}, \& {Thernoee}}]{stromgren44}
{Str{\"o}mgren}, B., {Gyldenkaerne}, K., {Rudkjobing}, M., \& {Thernoee}, K.~A.
  1944, Publikationer og mindre Meddeler fra Kobenhavns Observatorium, 138, 1

\bibitem[{{Tremblay} {et~al.}(2015){Tremblay}, {Gianninas}, {Kilic},
  {et~al.}}]{tremblay15}
{Tremblay}, P.-E., {Gianninas}, A., {Kilic}, M., {et~al.} 2015, \apj, 809, 148

\bibitem[{{Tremblay} {et~al.}(2013){Tremblay}, {Ludwig}, {Steffen}, \&
  {Freytag}}]{tremblay13}
{Tremblay}, P.-E., {Ludwig}, H.-G., {Steffen}, M., \& {Freytag}, B. 2013, \aap,
  559, A104

\bibitem[{{Vennes} {et~al.}(2011){Vennes}, {Thorstensen}, {Kawka},
  {et~al.}}]{vennes11}
{Vennes}, S., {Thorstensen}, J.~R., {Kawka}, A., {et~al.} 2011, \apjl, 737, L16

\bibitem[{{Webbink}(1984)}]{webbink84}
{Webbink}, R.~F. 1984, \apj, 277, 355

\bibitem[{{Zuckerman} {et~al.}(2007){Zuckerman}, {Koester}, {Melis}, {Hansen},
  \& {Jura}}]{zuckerman07}
{Zuckerman}, B., {Koester}, D., {Melis}, C., {Hansen}, B.~M., \& {Jura}, M.
  2007, \apj, 671, 872

\end{thebibliography}
\end{document}